\documentclass[iop]{emulateapj} 
\usepackage{graphicx}

\newcommand{\kms}{\,km\,s$^{-1}$}        \newcommand{\sqcm}{\,cm$^{-2}$}  
 \newcommand{\hi}{\ion{H}{1}}
\newcommand{\tm}{\tablenotemark}        \newcommand{\tn}{\tablenotetext}
\newcommand{\hst}{\emph{HST}}           \newcommand{\sw}{\ion{S}{2}}
\newcommand{\oi}{\ion{O}{1}}            
\newcommand{\sit}{\ion{Si}{3}}          \newcommand{\siw}{\ion{Si}{2}}
\newcommand{\nsl}{seven}                

\newcommand{\cds}{\object{CD14-A05}}
\newcommand{\uvqs}{\object{UVQS\,J1016-3150}}
\newcommand{\pks}{\object{PKS\,1101-325}}
\newcommand{\irasf}{\object{IRAS\,F09539-0439}}
\newcommand{\sdss}{\object{SDSS J0959+0503}}
\newcommand{\ngca}{\object{NGC\,3783}}
\newcommand{\ngcb}{\object{NGC\,3125}}

\begin{document}
\shorttitle{Chemical Abundances in the Leading Arm}
\shortauthors{Fox et al.}
\title{Chemical Abundances in the Leading Arm of the Magellanic 
Stream\footnotemark[*]}
\footnotetext[*]{Based on observations taken under programs
  12172, 12212, 12248, 12275, 13115, and 14687 
  of the NASA/ESA Hubble Space Telescope, obtained at the 
  Space Telescope Science Institute, which is operated by the Association of 
  Universities for Research in Astronomy, Inc., under NASA contract 
  NAS 5-26555, and under programs GBT12A\_206,
  GBT17B\_424 of the Green Bank Observatory,
  which is a facility of the National Science Foundation and is operated
  by Associated Universities, Inc.}
\author{Andrew J. Fox$^1$, Kathleen A. Barger$^2$, Bart P. Wakker$^3$, 
  Philipp Richter$^4$, Jacqueline Antwi-Danso$^{2,5}$,
  Dana I. Casetti-Dinescu$^6$, J. Christopher Howk$^7$, Nicolas Lehner$^7$,
  Elena D'Onghia$^{3,8}$, Paul A. Crowther$^9$, \& Felix J. Lockman$^{10}$}
\affil{$^1$ Space Telescope Science Institute, 3700 San Martin Drive,
  Baltimore, MD 21218\\
$^2$ Department of Physics and Astronomy, Texas Christian University, 
TCU Box 298840, Fort Worth, TX 76129\\
$^3$ Department of Astronomy, University of Wisconsin--Madison, 
475 North Charter St., Madison, WI 53706\\
$^4$ Institut f\"ur Physik und Astronomie, Universit\"at Potsdam, Haus
   28, Karl-Liebknecht-Str. 24/25, 14476, Potsdam, Germany\\
$^5$ Department of Physics \& Astronomy, Texas A\&M University, College Station,
   TX 77843\\
$^6$ Department of Physics, Southern Connecticut State University, 
501 Crescent Street, New Haven, CT 06515\\
$^7$ Department of Physics, University of Notre Dame, 225 Nieuwland 
Science Hall, Notre Dame, IN 46556\\
$^8$ Center for Computational Astrophysics, Flatiron Institute, 162 Fifth
Avenue, New York, NY 10010\\
$^9$ Department of Physics \& Astronomy, Hounsfield Road, University of 
Sheffield, S3 7RH, UK\\
$^{10}$ Green Bank Observatory, P.O. Box 2, Route 28/92, Green Bank, WV 24944}
\email{afox@stsci.edu}

\begin{abstract} 
The Leading Arm (LA) of the Magellanic Stream is a vast debris field
of \hi\ clouds connecting the Milky Way and the Magellanic Clouds.
It represents an example of active gas accretion onto the Galaxy. 
Previously only one chemical abundance measurement had been made in the LA.
Here we present chemical abundance measurements using 
\emph{Hubble Space Telescope}/Cosmic Origins Spectrograph 
and Green Bank Telescope spectra of four AGN sightlines passing through the LA
and three nearby sightlines 
that may trace outer fragments of the LA. 
We find low oxygen abundances, ranging from 4.0$^{+2.0}_{-2.0}$\% solar to 
12.6$^{+6.0}_{-4.1}$\% solar, in the confirmed LA directions, with the lowest values
found in the region known as LA III, farthest from the LMC.
These abundances are substantially lower than the single previous measurement,
S/H=35$\pm$7\% solar (Lu et al. 1998),
but are in agreement with those reported in
the SMC filament of the trailing Stream, 
supporting a common origin in the SMC (not the LMC)
for the majority of the LA and trailing Stream.
This provides important constraints for models of the formation
of the Magellanic System.
Finally, two of the three nearby sightlines show high-velocity clouds with
\hi\ columns, kinematics, and oxygen abundances
consistent with LA membership. This suggests that the LA is
larger than traditionally thought,
extending at least 20\degr\ further to the Galactic northwest.
\end{abstract}
\keywords{ISM: abundances -- Magellanic Clouds -- Galaxy: halo -- 
Galaxy: evolution -- quasars: absorption lines}

\section{Introduction}

Understanding the processes that deliver gas to galaxies is a major goal of 
modern galactic astrophysics. A key part of this effort is the observational 
characterization of gas clouds accreting onto galaxies. In the Milky Way, 
accreting gas can be seen directly
among the population of high velocity clouds
\citep[HVCs; see reviews by][]{WW97, Pu12, Ri17},
so there is a considerable body of knowledge
on the state of current-day Galactic accretion. Through a combination of 
radio 21 cm emission-line spectroscopy and UV absorption-line 
spectroscopy, the chemical abundances of HVCs can be measured. 
These abundances give important clues on their origin.

The Leading Arm (LA) of the Magellanic Stream is a well-known HVC complex and
a prime example of active accretion onto the Milky Way
\citep[see review by][]{DO16}. It forms a network of fragmented clouds 
connecting the LMC with the Galactic disk, and is visible in 
the original 21\,cm data that led to the discovery of the Magellanic Stream 
\citep{Wa72, Ma74}. Its connection to the Magellanic Clouds was 
demonstrated kinematically by \citet{Pu98}, and the fact that it \emph{leads}
the orbital motion of the Magellanic Clouds provides strong evidence for a
tidal origin, since ram pressure (the chief alternative mechanism
for formation of the Stream) cannot easily produce a leading arm.

The LA contains three principal substructures, named LA I, LA II, and LA III 
\citep{Pu98, Br05, For13}, though a fourth sub-structure
(LA IV) has been reported \citep{Ve12, For13}, and 
the full extent of the LA tidal debris field is not known.
The sub-structures likely lie at different distances from the Sun.
Region LA I lies below the Galactic plane at
$d\!\approx\!20$\,kpc, based on the detection of a young 
stellar population \citep{CD14, Zh17}.
Regions LA II and LA III lie above the plane; 
the presence of many cometary head-tail clouds \citep{For13}
in between LA II and LA III suggests that this
inter-cloud region has already reached (and is interacting with)
the outer disk of the Milky Way at a distance of
$\approx$21~kpc \citep{MG08, CD14}. 

Previous studies of the chemical enrichment of the LA are limited to a single 
sightline, toward \ngca\ \citep{Lu94, Lu98, Se01}, which passes
through LA II. \citet{Lu98} analyzed \emph{Hubble Space Telescope}/Goddard
High Resolution Spectrograph (\hst/GHRS) spectra 
and derived a LA sulfur abundance (S/H)=0.35$\pm$0.07 solar 
and a sulfur-to-iron ratio (S/Fe)=10.7$\pm$2.2 solar, indicative of
depletion of iron atoms into dust grains
[these numbers have been updated to reflect the \citet{As09} solar abundances,
rather than the \citet{AG89} solar abundances used by Lu et al.].
The quarter-solar metallicity
contrasts with measurements in the Magellanic Stream (MS), where
a metallicity of $\approx$0.1 solar or lower has been measured in
seven directions \citep{Fo10, Fo13, Ku15, Ho17}, supporting an SMC origin,
although there is an LMC filament of the Stream 
with much higher metallicity \citep[$\approx$0.5 solar;][]{Gi00, Ri13}
and different kinematics \citep{Ni08}.
This indicates that the Stream has a dual origin.

So where and when did the LA originate? Did its origin coincide with
the formation of the Magellanic Stream, as tidal models predict?
These are the fundamental questions
addressed in this paper. We present new and archival \hst/Cosmic Origins
Spectrograph (COS) spectra of four sightlines passing through the LA, and
three sightlines passing nearby, together with \hi\ 21 cm
observations of the same directions from existing radio surveys
and from new observations.
We analyze the \oi/\hi\ and \sw/\hi\ ratios to determine the
chemical abundances in regions LA I, LA II, and LA III, which are at widely
different angular scales from the LMC. We also investigate whether
the HVCs seen in the three nearby sightlines could be physically associated with
the LA, to explore its total angular size.

\section{Observations and Data Handling}
\label{sec:observations}

\begin{deluxetable*}{lllcc llccc}
\tabletypesize{\small}
\label{tab:sample}
\tabcolsep=2.0pt
\tablewidth{0pt}
\tablecaption{The Sample: {\it HST}/COS Sightlines through and near the Leading Arm}
\tablehead{Target & Type & Region\tm{a} & $l$     & $b$        & Program ID\tm{b} & Grating & $v_0$(\hi)\tm{c} & log\,$N$(\hi)\tm{c} & Source\tm{d}\\
                  &      &              & (\degr) & (\degr)    & & & (km\,s$^{-1}$) & ($N$ in cm$^{-2}$)}
\startdata
\multicolumn{2}{l}{\bf Confirmed LA Sightlines\tm{e}}\\
CD14-A05          & B5IV    & LA I    & 295.7476 & $-$11.7813 & 14687 & G130M+G160M & +120 & 18.83$\pm$0.10\tm{f} & GASS\\
NGC\,3783         & Sey1    & LA II   & 287.4560 & +22.9476   & 12212,13115 & G130M+G160M & +240 & 19.92$\pm$0.10 & ATCA\\
NGC\,3125 & \ion{H}{2} Gal. & LA III  & 265.3254 & +20.6448   & 12172 & G130M       & +210 & 19.05$\pm$0.05 & GBT\\ 
UVQS\,J1016-3150\tm{g} & QSO & LA III  & 268.3692 & +20.4379   & 14687 & G130M      & +210 & 18.47$\pm$0.05 & GBT\\ 
\hline
\multicolumn{2}{l}{\bf Potential LA Sightlines\tm{e}}\\
PKS\,1101-325      & Sey1    & LA II/III\tm{h} & 278.1171 & +24.7652 & 12275 & G130M& +120 & 19.30$\pm$0.05 & GBT\\ 
IRAS\,F09539-0439  & Sey1    & LA Ext. & 243.3317 & +37.0045  & 12275 & G130M       & +160 & 19.21$\pm$0.05 & GBT\\ 
SDSS\,J0959+0503\tm{i} & QSO & LA Ext. & 233.3699 & +43.4836  & 12248 & G130M+G160M & +289 & 18.74$\pm$0.10 & EBHIS
\enddata
\tn{a}{Regions of Leading Arm are defined as in \citet{For13}, except LA Ext. (Extension) defined here.}
\tn{b}{Program ID of \hst/COS dataset.}
\tn{c}{Central velocity and column density of \hi\ 21 cm emission from LA component.}
\tn{d}{Telescope or survey used for radio data: GASS \citep{MG09}, EBHIS \citep{Wi16}, GBT (this paper) or ATCA \citep{Wa02}.}
\tn{e}{Confirmed LA sightlines are those passing through the 21 cm emission contours from regions LA I, LA II, LA III, plus (in the case of CD14-A05) those to confirmed LA stellar sources. Potential LA sightlines are nearby and show high-velocity 21\,cm emission but their physical connection to the LA is unconfirmed.}
\tn{f}{Some of this \hi\ column may lie behind the star, which lies at $d\approx$18--20 kpc \citep{Zh17}.} 
\tn{g}{Full name UVQS J101629.20-315023.6. We use an abbreviated name for brevity.}
\tn{h}{This sightline lies midway between regions LA II and LA III, where a bridge of \hi\ clumps is reported \citep{For13}.}
\tn{i}{Full name SDSS J095915.60+050355.0. We use an abbreviated name for brevity.}

\end{deluxetable*}


\begin{figure*}[!ht]
\epsscale{1.0}
\label{fig:lamap}

\plotone{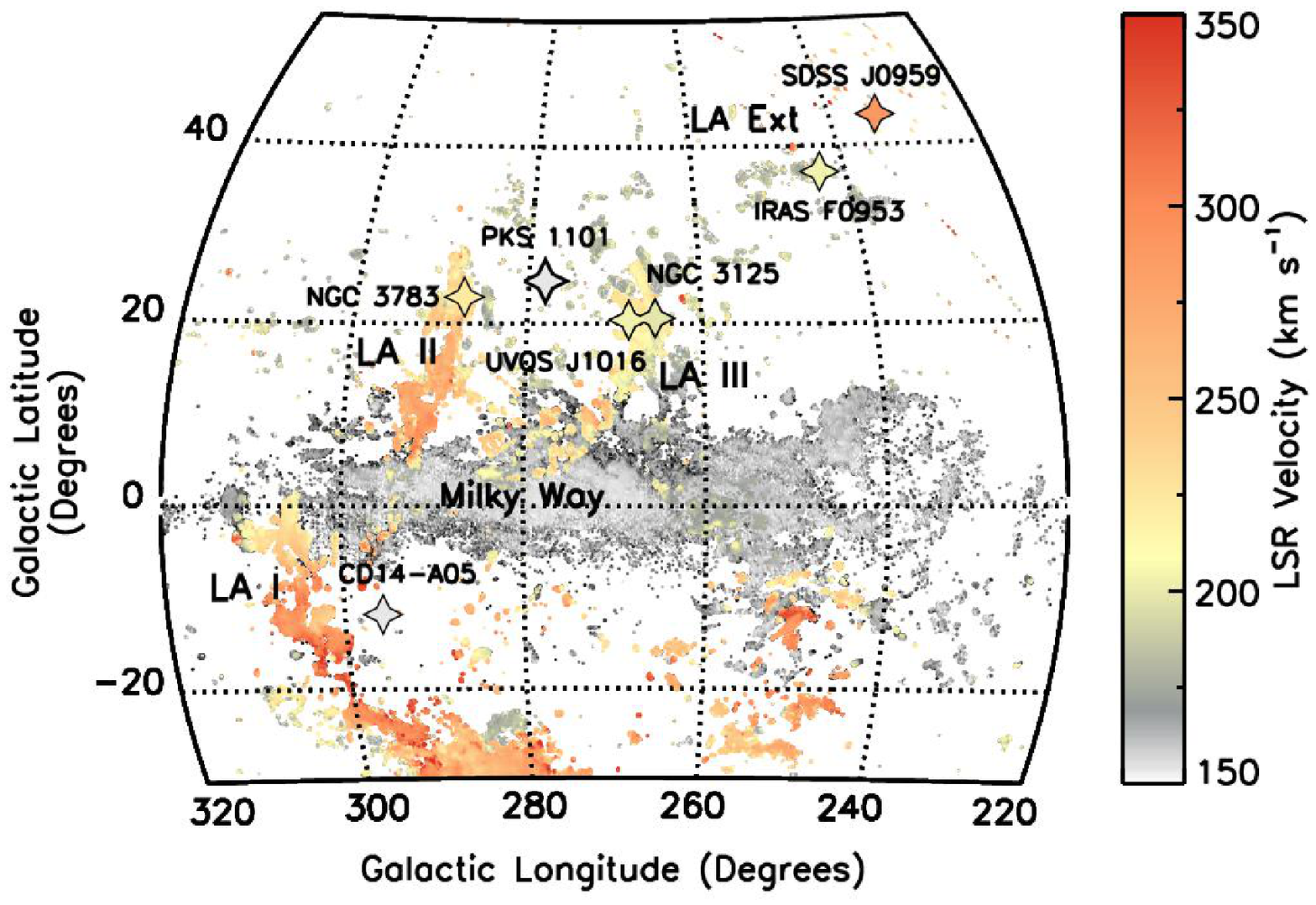} 
\plotone{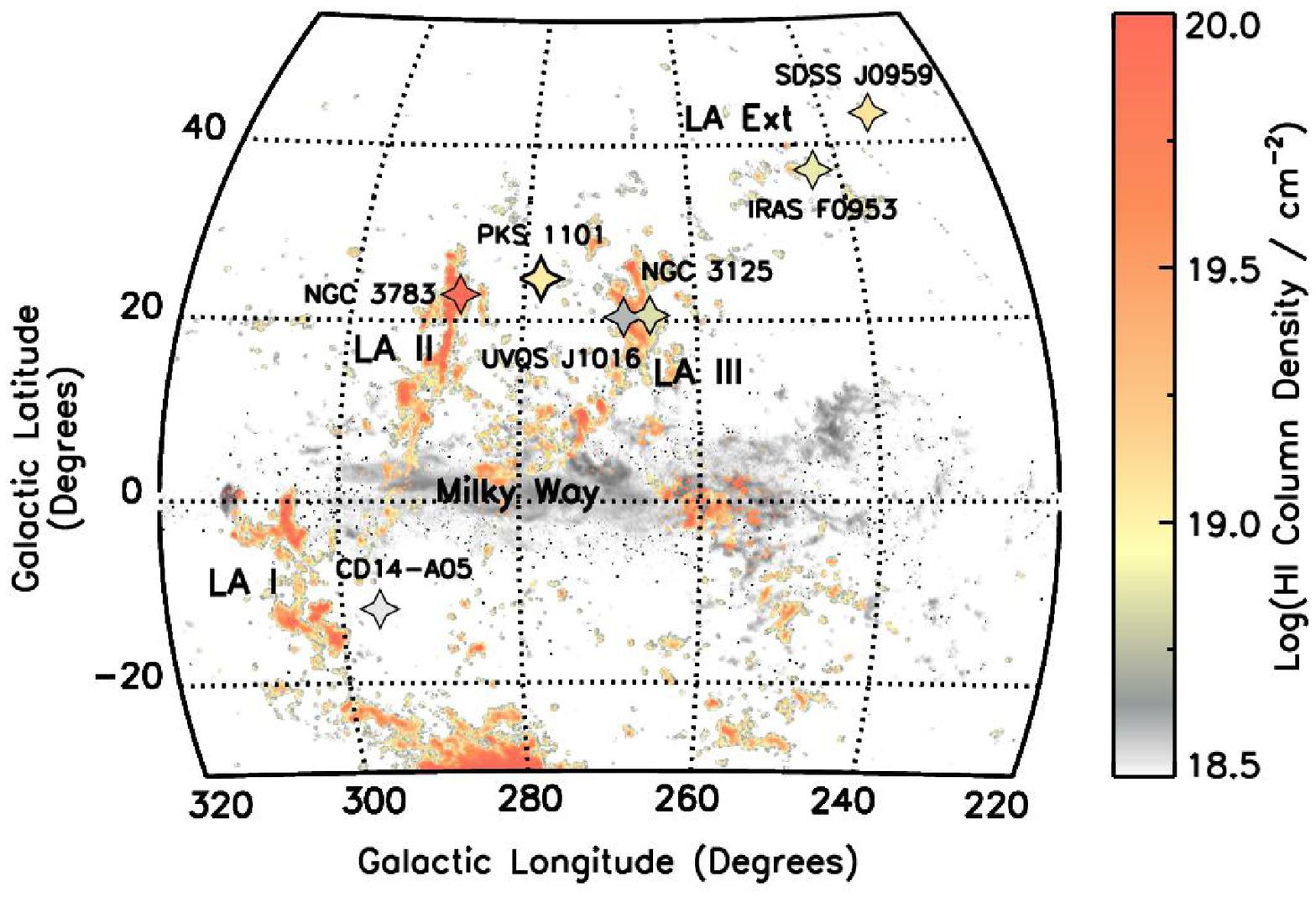} 
\caption{
  Maps of the LA showing the location of our \hst/COS sightlines (stars)
  relative to the \hi\ emission from the GASS survey \citep{MG09}.
  The upper map shows the \hi\ velocity field with the COS sightlines
  color-coded by velocity of absorption. The lower map shows the
  \hi\ column densities with the COS sightlines color-coded by $N$(\hi).
  The \hi\ data in the top-right corner of each map
  ($l\!<\!240$\degr, $b\!>\!40$\degr) are taken from the LAB survey
  \citep{Ka05} since GASS does not cover this region.
  Both maps show \hi\ data in the range $150\!<\!v_{\rm LSR}\!<\!350$\kms,
  so the Galactic foreground is strong at low latitude. 
The scattered morphology and complex velocity field of the LA are clear.
The CD14-A05 sightline is slightly offset from region LA I,
but this star has confirmed kinematic membership of the LA \citep{CD14, Zh17}
and the absorption velocity matches the stellar velocity.}
\end{figure*}

\subsection{Sample Creation}
We were awarded six orbits of \hst/COS time in Cycle 24 under Program ID 
14687 to observe two LA targets:
(1) the B5IV star \cds\ lying close to region LA I; this star
was first identified as a Magellanic System candidate by \citet{CD12},
has a photospheric metallicity [Mg/H]=$-$0.57$\pm$0.35,
and a radial velocity $v_{\rm r}$=133$\pm$9\kms\ \citep{Zh17};
(2) the QSO \uvqs\ at $z$=0.2417 \citep[identified as a QSO by][]{Mo16}
lying behind LA III.
Both these targets show \hi\ 21 cm detections at LA 
velocities, making them suitable for abundance analyses.
We also searched the \hst\ archive for COS spectra of
UV-bright AGN lying either behind or projected near to the LA,
and uncovered five more sightlines with \hi\ 21 cm
detections from the LA, making a total sample of \nsl\
sightlines\footnotemark[1] (see Table 1). 
\footnotetext[1]{In \citet{Fo14}, we presented 16 COS spectra of
sources in the Leading Arm region as part of a survey of 69 sightlines 
through the extended Magellanic System. However, only three 
of these 16 have \hi\ detections, and those are included in the
current sample.}

The \nsl\ directions in our sample are divided into two categories.
The first category
are the four \emph{Confirmed LA Sightlines}. These pass through the
known \hi\ regions LA I, LA II, or LA III, or (in the case of \cds)
are confirmed since the target itself is a member of the LA stellar
population \citep{CD14, Zh17}.
The second category are the three \emph{Potential LA Sightlines}.
These are directions that lie within $\approx$30\degr\ of the
main LA regions and which show 21 cm \hi\ emission, but
have an unconfirmed physical connection to the LA.
Two of these sightlines pass through the \hi\ debris field
that lies $\approx$15--30$\degr$ northwest of region LA III.
The third (\pks) lies directly in-between LA II and LA III.
The HVCs observed in these directions may be physically associated with
the LA, and we use their chemical properties and kinematics to explore
this possibility.

The location of each direction with respect to the \hi\ emission is shown in
Figure~1, where we include both column density maps and velocity field
maps (zeroth and first moment maps). 
The lack of \hst\ targets at low latitude is due
to extinction from the Galactic disk, preventing background UV sources
from being detectable, at least with current instrumentation. This precludes the
measurement of LA abundances in the latitude range $-10\degr\la b\la10\degr$,
where the LA crosses (and potentially interacts with) the Galactic disk.

\subsection{Data Reduction}
For each COS target, we took the extracted one-dimensional spectra 
from the {\tt calcos} 
pipeline ({\tt x1d} files) and aligned them in wavelength space using 
customized reduction software that cross-correlates the positions
of low-ionization ISM lines \citep[following][]{Wa15}.
The aligned spectra are co-added and then further calibrated
using the velocities of interstellar 21\,cm \hi\ components as zero-points.
Lines due to intergalactic systems at higher redshift were identified.
The data were continuum-normalized around each absorption line of interest.
The spectra were binned by five pixels (to 10\kms\ bins, to give two rebinned
pixels per resolution element) for display purposes. 
A second, night-only reduction was conducted to extract the data 
taken during orbital nighttime. This reduces geocoronal airglow emission
in the range $-100\!\la\!v_{\rm LSR}\!\la\!100$\kms, which 
allows us to measure the \oi\ 1302 line crucial to our abundance analysis.
The spectral resolution of the COS data is $R\!\approx$15000--18000
(FWHM$\approx$17--20\kms), depending on wavelength and detector lifetime position.

A stack of absorption lines for each direction is presented in Figure 2.
The suite of lines shown
depends on whether the target was observed with both the G130M and G160M
gratings (three targets, with coverage from $\approx$1150--1700\AA), or
just the G130M grating (four targets, with coverage from
$\approx$1150--1450\AA).
For the sightlines observed with both gratings, we show
\oi\ $\lambda$1302 (night-only data), \ion{Al}{2} $\lambda$1670,
\ion{Si}{2} $\lambda\lambda$1260, 1193, 1190, 1526,
\ion{Si}{3} $\lambda$1206, \sw\ $\lambda\lambda$1250, 1253,  
\ion{Fe}{2} $\lambda\lambda$1144, 1608,
\ion{C}{4} $\lambda\lambda$1548, 1550, and
\ion{Si}{4} $\lambda\lambda$1393, 1402. 
These lines were chosen as the strongest metal-lines available in the COS 
far-ultraviolet (FUV) bandpass. We do not show \ion{C}{2} $\lambda$1334
or \sw\ 1259 since these lines are blended at LA velocities
by Galactic \ion{C}{2}$^*$ 1335 and \ion{Si}{2} 1260, respectively.
Atomic data were taken from \citet{Mo03} and verified against the recent
updates by \citet{Ca17}.

\begin{figure*}[!ht]
\label{fig:stacks}
\setcounter{figure}{1}
\epsscale{1.1}
\plotone{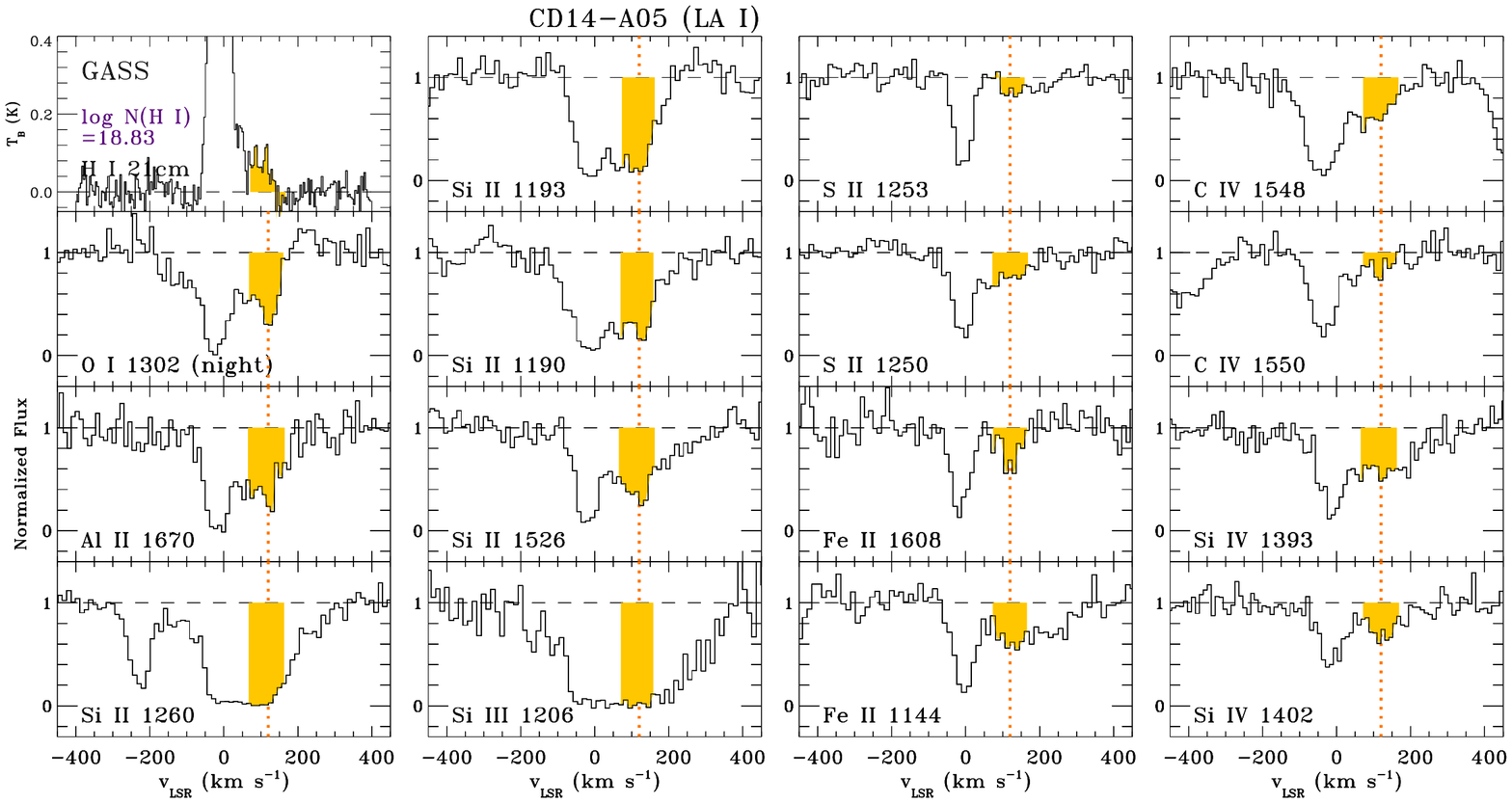} 
\plotone{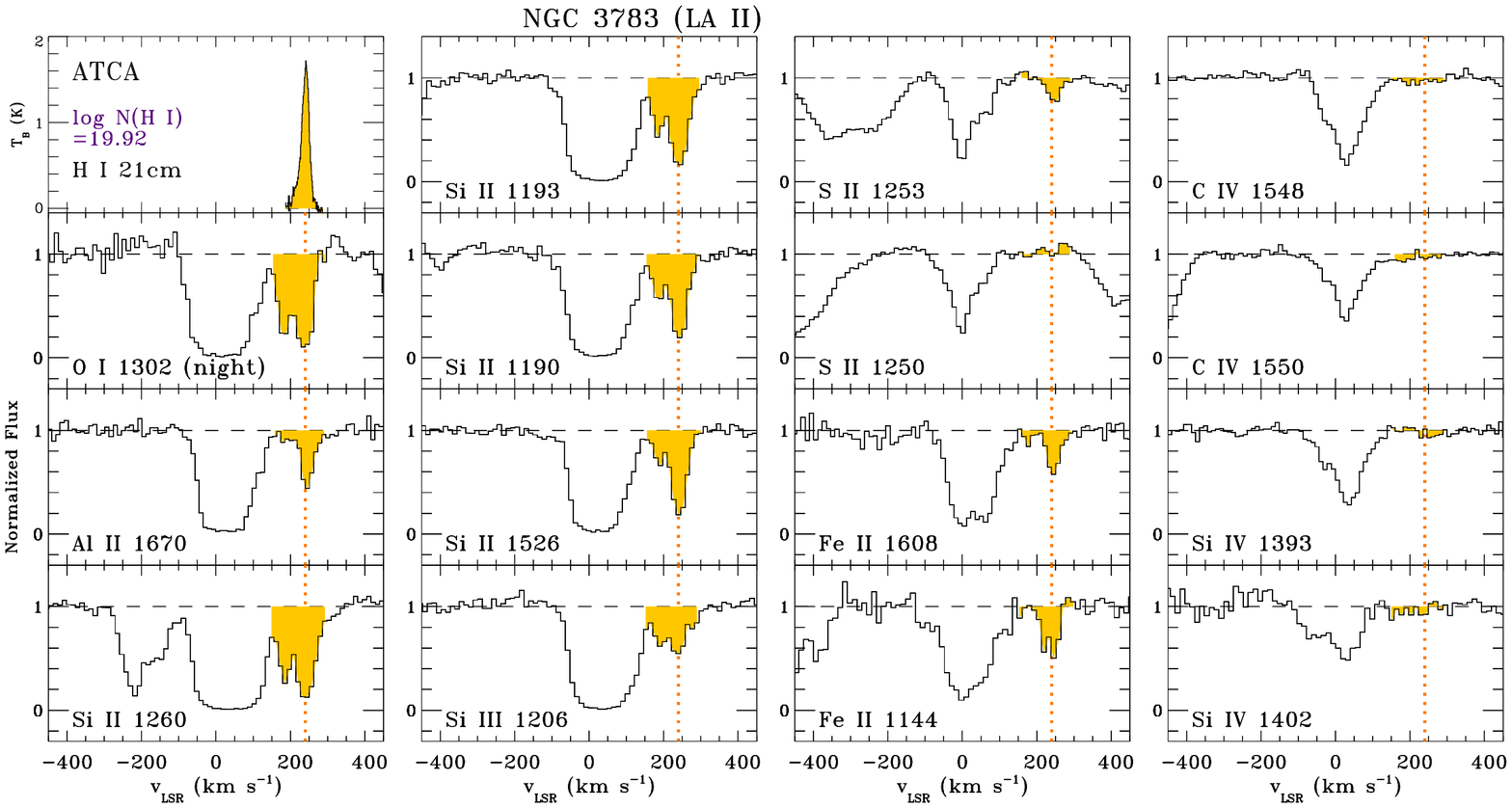}
\plotone{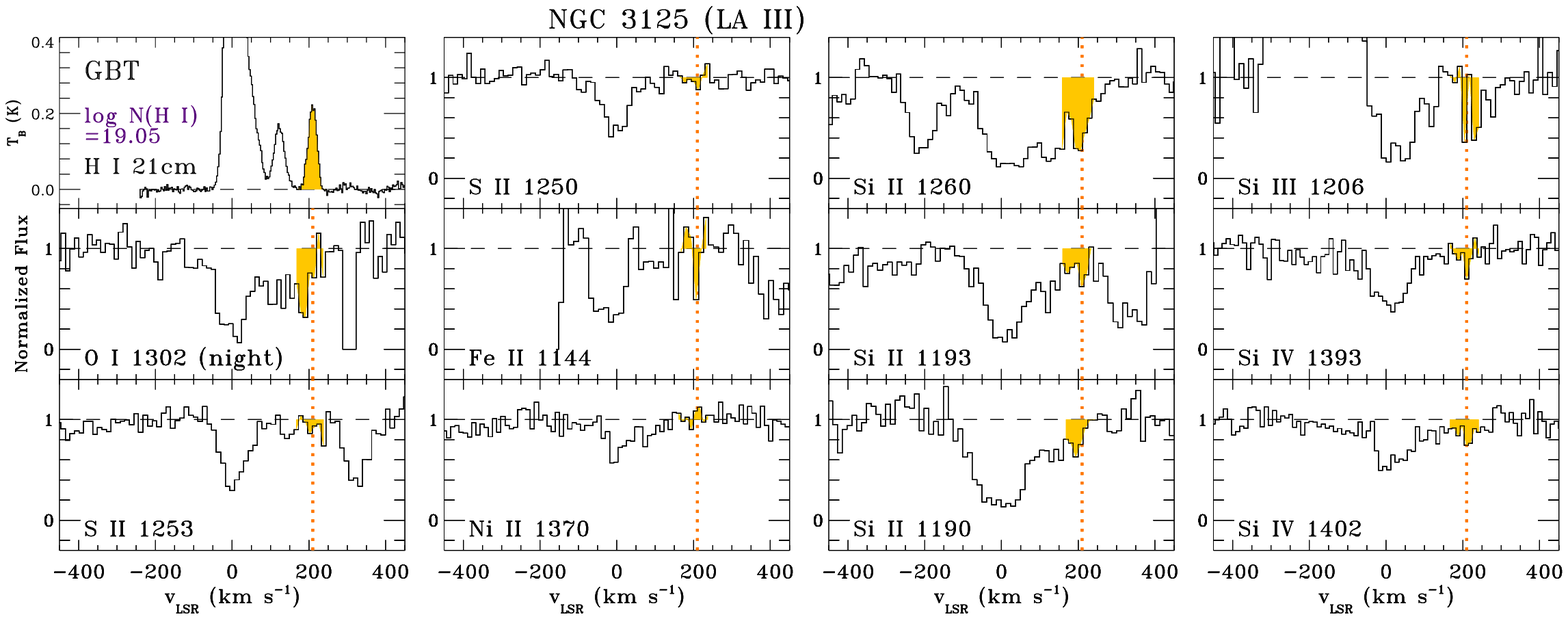}
\caption{\hst/COS spectra of UV metal-line absorption in
  each sightline in our sample. Normalized flux is plotted against LSR
  velocity for a range of low-ion and high-ion transitions. In each direction,
  a 21 cm \hi\ emission profile is included in the top-left panel. Golden
  shading denotes the LA component; the LA velocity centroid is indicated by the
  vertical dashed line.
  The COS data and the \hi\ data have been rebinned by five pixels for display purposes.
  The \oi\ $\lambda$1302 data are from a night-only reduction.}
\end{figure*}

\begin{figure*}[!ht]
\setcounter{figure}{1}
\epsscale{1.1}
\plotone{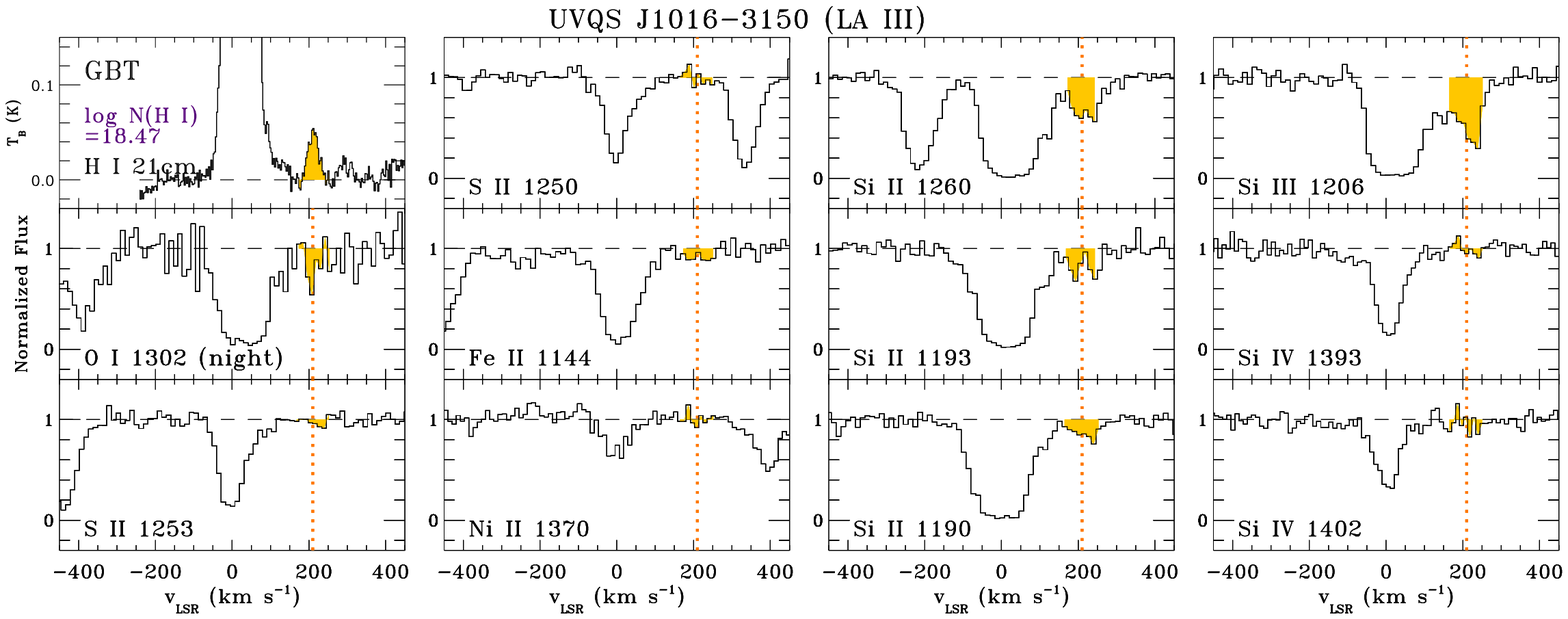}
\plotone{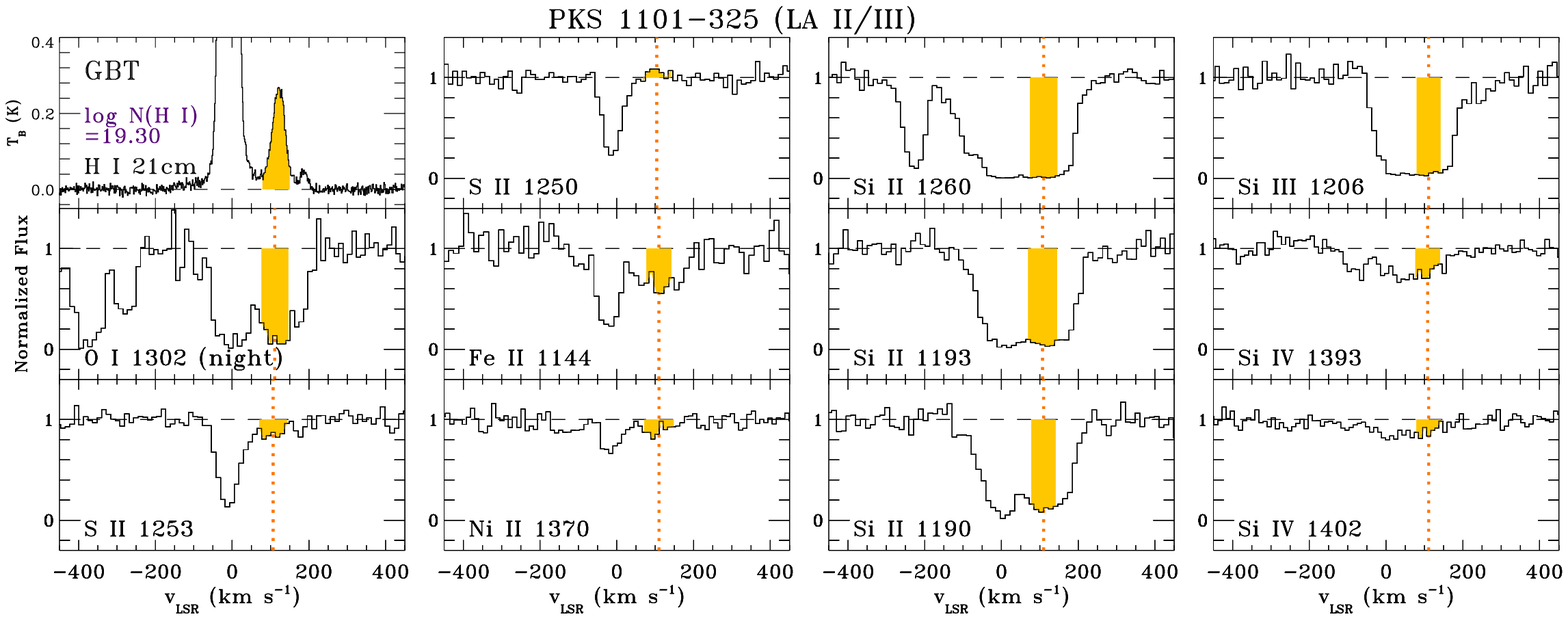}
\caption{(cont). \hst/COS and 21\,cm \hi\ spectra of each sightline in our sample.}
\end{figure*}

\begin{figure*}[!ht]
\setcounter{figure}{1}
\epsscale{1.1}
\plotone{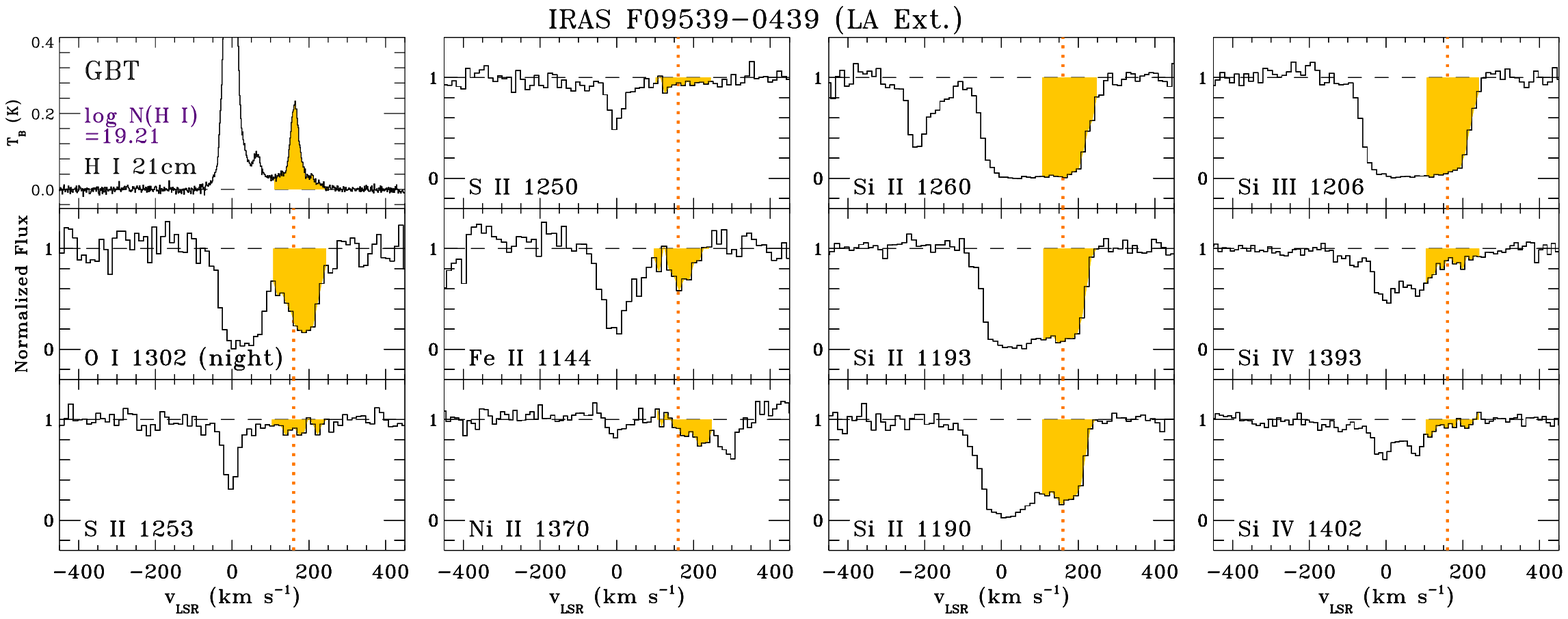}
\plotone{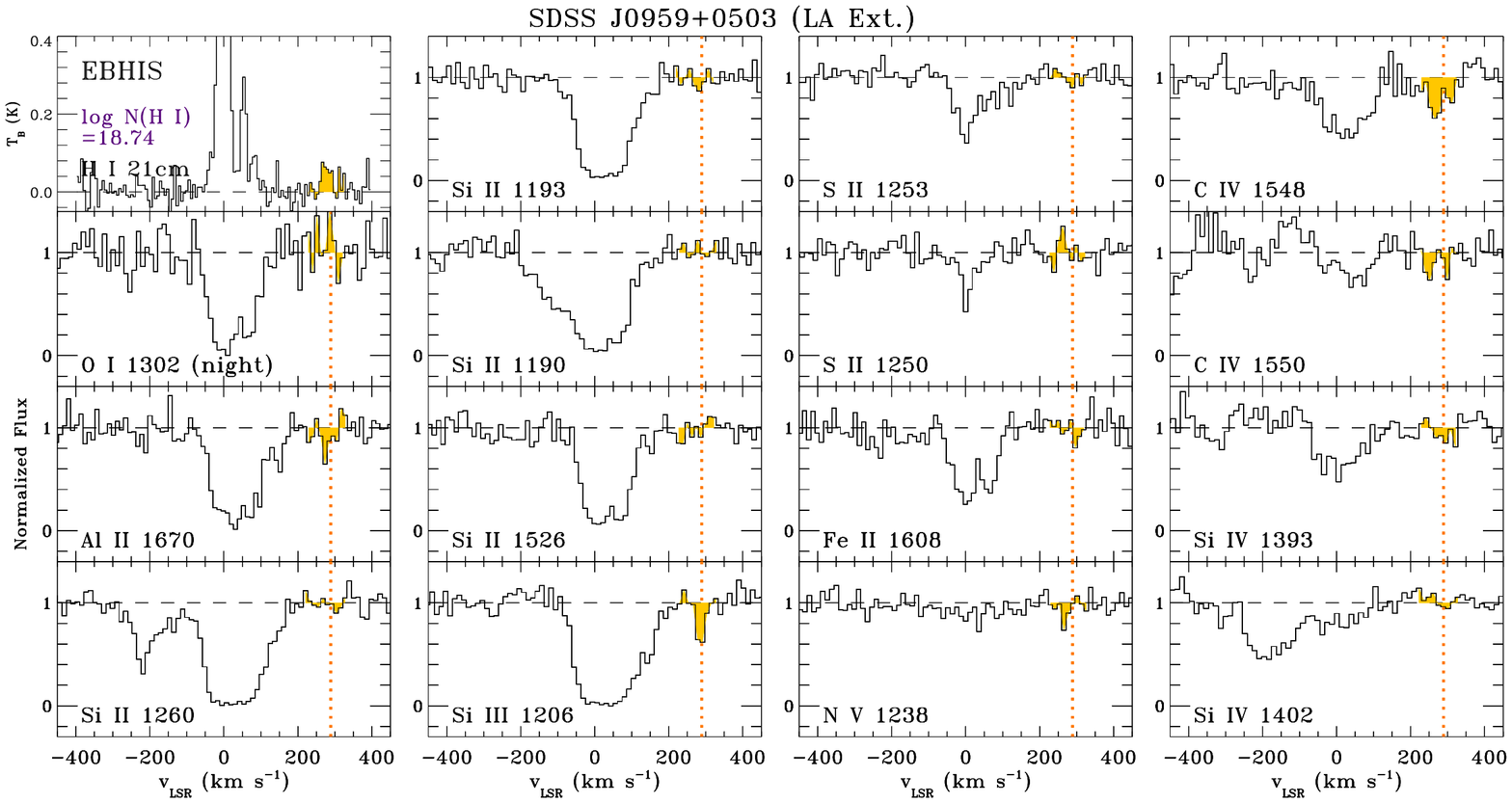} 
\caption{(cont). \hst/COS and 21\,cm \hi\ spectra of each sightline in our sample.}
\end{figure*}

\subsection{21 cm \hi\ Spectra}
For the reference \hi\ column densities, we use 21 cm spectra from a
variety of telescopes, as summarized in Table 1.
For four of our directions
we obtained new high-sensitivity observations from the Robert C. Byrd Green
Bank Telescope (GBT).
These data were taken under programs GBT12A\_206 and GBT17B\_424.
The spectra were measured over the LSR velocity range $-$450 and +550\kms\
at a velocity resolution of 0.15\kms\ using frequency switching. 
The spectra were smoothed, converted to brightness temperature, and corrected
for stray radiation following the procedures described in \citet{Bo11}.
A third-order polynomial was fit to emission-free regions of the
spectra to achieve a final rms noise in brightness temperature of $\approx$11\,mK
in a 0.6 \kms\ channel.
The GBT has an angular resolution of 9.1\arcmin\ at 1420 MHz.

For the other three directions, we use \hi\ 21 cm spectra from either
(1) the Galactic All-Sky Survey \citep[GASS; 16.1$\arcmin$ angular
resolution;][]{MG09},
(2) the Effelsberg-Bonn \hi\ Survey 
\citep[EBHIS; 10.8$\arcmin$ resolution;][]{Wi16}, or
(3) the Australia Telescope Compact Array \citep[ATCA; 1\arcmin-resolution;][]{Wa02}.
The choice of dataset adopted in each direction depends on the target declination
and availability of data;
we present the highest angular resolution spectra available for each target.

\subsection{Measurement of Absorption}
\label{sec:measurement}

We use the known 21 cm velocity field of the LA (shown in Figure~1a)
to identify LA absorption components.
The identification step is important since non-Magellanic absorption 
components (from foreground or background structures)
could be present in the data at other velocities, and without the 21 cm
velocity field one could misidentify them as LA components.

The \hi\ velocity field in the LA is complex,
extending from below $\approx$150\kms\ to $\approx$350\kms\ (Fig. 1a).
The traditionally-defined regions LA I, LA II, and LA III lie in the
higher-velocity end ($\approx$200--350\kms) of this range,
though gas at $\approx$150\kms\ is found
near each of LA I, LA II, and LA III.
For the seven sightlines in our survey, the LSR velocities of the LA
components we identify range from +120 to +289\kms.
For context, the velocity field for high-positive-velocity gas in this quadrant
of the sky is given in \citet{Rich17}. Their Figure~8 shows an extended
population of high-positive-velocity UV absorbers (halo clouds) extending to
the Galactic
north-west beyond the boundaries of the LA, in the direction of Complexes
WA and Complex M, but without 21 cm counterparts. Therefore there are no
known 21\,cm HVCs other than the LA in this part of the sky.



The two ions we focus on for chemical abundances measurements are \sw\ and \oi,
traced via \sw\ $\lambda\lambda$1250, 1253 (the $\lambda$1259 region
is blended) and 
\oi\ $\lambda$1302. These ions are the best metallicity indicators
available in the FUV because sulfur and oxygen are relatively undepleted
onto dust grains \citep{Je05, Je09}
and have relatively small ionization corrections
(particularly for oxygen; see Section 3.1). 
They are also both $\alpha$-elements so have similar nucleosynthetic origins.
Column densities for these two ions were determined via the apparent
optical depth
(AOD) technique \citep{SS91}, which returns accurate measures of the 
column density provided the line profiles are resolved and unsaturated.
The AOD in each pixel is given by $\tau_a(v)$=ln\,$[F_c(v)/F(v)]$,
where $F(v)$ and $F_c(v)$ are the observed flux and the estimated
continuum flux, respectively, as a function of velocity. The integrated AOD is 
$\tau_a=\int_{v_{\rm min}}^{v_{\rm max}}\tau_a(v){\mathrm d}v$, and
the apparent column density is
$N_{\rm a}(v)=3.768\times10^{14}(f\lambda)^{-1}\tau_a(v)$\sqcm.
The velocity range of absorption, $v_{\rm min}$ to $v_{\rm max}$,
is selected by eye to encompass the range 
of the UV metal-line absorption and the \hi\ emission, though we include a
contribution to the error budget to account for varying these limits by $\pm$5\kms.

The AOD measurements in each direction are presented in Table 2.
In lines that are clearly or potentially saturated
(those with normalized flux $F(v)/F_c(v)\!<\!0.1$)
we present a lower limit on $N_a(v)$. In lines that are not detected
at 3$\sigma$
significance, we present an upper limit on $N(v)$, based on measuring
a 3$\sigma$ limit on the equivalent width and converting it to column density
assuming a linear curve-of-growth.

\begin{deluxetable*}{lcccc ccc}
\tabletypesize{\small}
\label{tab:meas}  
\tabcolsep=2.0pt
\tablewidth{0pt}
\tablecaption{Measurements of Leading Arm Absorption and Emission}
\tablehead{Target & $v_{\rm min}$\tm{a} & $v_{\rm max}$\tm{a} & log\,$N$(\hi)\tm{b} & log\,$N$(\sw)\tm{c} & log\,$N$(\oi)\tm{d} & [\sw/\hi]\tm{e} & [\oi/\hi]\tm{f}\\
             & (\kms) & (\kms) & ($N$ in cm$^{-2}$) & ($N$ in cm$^{-2}$) & ($N$ in cm$^{-2}$) & & }
\startdata
\multicolumn{2}{l}{\bf Confirmed LA Sightlines}\\
CD14-A05      & $\phn$75 & 170 & 18.83$\pm$0.10 & 14.46$\pm$0.22\tm{g} & 14.64$\pm$0.10 & +0.51$\pm$0.26 & $-$0.90$\pm$0.17 \\
NGC\,3783          & 160 & 300 & 19.92$\pm$0.10 & 14.41$\pm$0.12 & $>$14.95\tm{h} & $-$0.63$\pm$0.16 & $>-$1.67\tm{h} \\ 
NGC\,3125          & 170 & 245 & 19.05$\pm$0.05 & $<$14.90       & 14.37$\pm$0.28 & $<$+0.73         & $-$1.40$\pm$0.30 \\
UVQS J1016-3150    & 175 & 255 & 18.47$\pm$0.05 & $<$14.62       & 13.96$\pm$0.31 & $<$+1.03         & $-$1.21$\pm$0.33\\
\hline
\multicolumn{2}{l}{\bf Potential LA Sightlines}\\
PKS\,1101-325 & $\phn$80 & 150 & 19.30$\pm$0.05 & 14.52$\pm$0.17 & $>$15.04            & +0.10$\pm$0.20   & ... \\
IRAS\,F09539-0439  & 110 & 250 & 19.21$\pm$0.05 & $<$14.93       & 14.94$\pm$0.09 & $<$+0.60         & $-$0.99$\pm$0.14\\
SDSS\,J0959+0503   & 230 & 330 & 18.74$\pm$0.10 & $<$14.82       & $<$14.54       & $<$+0.96         & $<-$0.93  \\
\vspace{-4mm}
\enddata
\tn{a}{Minimum and maximum LSR velocities of LA emission and absorption.}
\tn{b}{\hi\ column is integrated using $N$(\hi)=$1.823\times10^{18}{\rm cm}^{-2}\int^{v_{\rm max}}_{v_{\rm min}} T_B\,{\rm d}v$.}
\tn{c}{Column integrated by AOD technique \citep{SS91}. \sw\ columns derived from 1253 line. Upper limits are 3$\sigma$ (non-detections).}
\tn{d}{\oi\ 1302 measurement is made on night-only data, to remove geocoronal emission. Lower limits given for saturated lines.}
\tn{e}{[\sw/\hi]=[log\,$N$(\sw)--log\,$N$(\hi)]--(S/H)$_\odot$. Beam-smearing error of 0.10\,dex has been added in quadrature with the statistical error on the \sw\ and \hi\ column densities.}
\tn{f}{[\oi/\hi]=[log\,$N$(\oi)--log\,$N$(\hi)]--(O/H)$_\odot$. Beam-smearing error of 0.10\,dex has been added in quadrature with the statistical errors.}
\tn{g}{AOD profile analysis indicates this line may be contaminated by an unidentified blend.}
\tn{h}{Unresolved saturation possible in \oi\ $\lambda$1302 so $N$(\oi) and [\oi/\hi] are lower limits. See Richter et al. (2018) for further analysis.}
\end{deluxetable*}


\section{Abundance Determinations}
\label{abundances}

Following standard notation, the ion abundances of \oi\ and \sw\ are defined on
a logarithmic scale relative to solar:\\

[\oi/\hi]=[log\,$N$(\oi)--log\,$N$(\hi)]--log\,(O/H)$_\odot$ and

[\sw/\hi]=[log\,$N$(\sw)--log\,$N$(\hi)]--log\,(S/H)$_\odot$.\\

We use the solar oxygen and sulfur abundances of \citet{As09}, which are
log\,(O/H)$_\odot$=--3.31 and log\,(S/H)$_\odot$=--4.88. The ion abundances
simply represent the observed concentration of ions relative to hydrogen,
without any ionization correction being applied.


\subsection{Ionization Corrections}
\label{ioncorrs}
Due to differences in ionization potentials (IPs) between various ions and \hi,
ionization effects can lead to differences between the ion abundances and the
true elemental abundances. These differences are known as ionization
corrections (ICs), where\\

[O/H]=[\oi/\hi]+IC(O) and

[S/H]=[\sw/\hi]+IC(S).\\

The magnitude of the IC depends on the element. For oxygen, the close similarity
in the first IPs of hydrogen (13.60\,eV) and oxygen (13.62\,eV) coupled with
charge-exchange reactions make the ICs very small \citep{Vi95, FS71},
except in cases of intense radiation fields.
For sulfur, the (larger) IP of \sw\ (23.34\,eV) means the ion can exist
in regions where the hydrogen is largely ionized, and so the IC is larger.

To calculate the ICs, we computed a grid of photoionization simulations using
the \emph{Cloudy} radiative transfer code \citep{Fe13}.
The simulations use a plane-parallel geometry and assume the plasma has
uniform density, so do not account for clumpiness in the gas.
For a given set of input parameters, \emph{Cloudy} calculates the ionization
breakdown of all chemical elements up to zinc.
The ICs are a function of the ionization parameter
$U \equiv n_\gamma/n_{\rm H}$ (the ratio of the ionizing photon density
to the gas density) and the \hi\ column density $N$(\hi) in the LA component.
These parameters are constrained as follows:\\

(1) To derive $n_\gamma$, we use a 3-D model of the Galactic ionizing
radiation field from \citet{Fo14}, which is based on \citet{BM99} and
updated by \citet{Ba13} to include the radiation from the Magellanic Clouds.
This is combined with the \citet{HM01} model of the extragalactic radiation
field at $z$=0. The LA is taken to be at $d$=20\,kpc, based on the
existing distance constraints towards LA I \citep{CD14, Zh17},
though see Section 3.2 for a discussion of the effect of changing the distance.\\

(2) $U$ is derived from the observed \sit/\siw\
column-density ratio in the LA; this ratio is 
a monotonic function of $U$ for various ionizing radiation fields
\citep{Fo16, Bo17}. Knowing $U$ and $n_\gamma$ allows $n_{\rm H}$ to be calculated.
LA directions where both \sit\ $\lambda$1206
and one of the \siw\ lines ($\lambda$1260, 1190, 1193, 1190)
are unsaturated (and thus measurable) have the best constrained models.
When \sit\ is saturated, only a lower limit on the \sit/\siw\ ratio
can be measured,
which translates to an lower limit on log\,$U$ and a limit on IC(S).
Fortunately IC(O) is independent of log\,$U$ when log\,$N$(\hi)$>$18
\citep{Bo17}, as is the case for all seven sightlines in our sample.
Therefore uncertainty in the knowledge of log\,$U$ does not affect the reliability
of our oxygen abundances.\\

(3) $N$(\hi) is directly measured in the 21\,cm data by integrating
over the velocity range of the LA component.\\

By running \emph{Cloudy} models in this manner, we derived the ICs and 
the corrected sulfur and oxygen abundances presented in Table~3.
Notice that for the range of log\,$N$(\hi) in our LA sample (18.47 to 19.92), 
the ICs for oxygen are negligible ($|$IC(O)$|\la0.04$\,dex) 
but the ICs for sulfur are substantial (up to $-$0.77\,dex), 
particularly in the cases where log\,$N$(\hi)$<$19.0.
The oxygen abundances should therefore be considered more robust.

\subsection{Uncertainties in the Abundance Measurements}
We investigated the uncertainty in the abundance measurements caused by six sources
of error: the distance to the LA, the velocity integration range,
the beam-size mismatch between radio and UV observations, the error on log\,$U$, 
the statistical error in the measurements of UV absorption,
and the statistical error in the measurements of \hi\ emission.
The first four of these are now discussed in turn (the last two are self-explanatory).
The six sources of error were added in quadrature to produce the final errors on
the abundance measurements presented in Table 3.
The statistical errors on the UV absorption measurements dominate the errors.\\

(1) {\it Uncertainty in distance}: to investigate the uncertainty in the ICs caused by
uncertainty in the distance to the LA, we ran {\it Cloudy} models for the NGC\,3125
sightline for five heliocentric distances: 10, 20, 30, 40, and 50 kpc, where
20\,kpc is the nominal case becuase of existing distance constraints.
This sightline (through LA III) was chosen to be representative of
the LA directions, given its intermediate value of log\,$N$(\hi)=19.05.
All other properties of the models were held constant;
increasing the distance decreases the Galactic contribution to the ionizing radiation field
while keeping the UV background contribution fixed.
For each distance, we computed the best-fit value of log\,$U$, IC(S), and IC(O).
We found that 
IC(S) varies from $-$0.42 at 10\,kpc to $-$0.35 at 50\,kpc, whereas
IC(O) remains flat at $-$0.01 at all distances. The distance error is therefore
negligible for the oxygen abundances and reaches $-$0.07\,dex for the sulfur abundances.\\

(2) {\it Uncertainty in velocity integration range}:
in each direction the velocity integration range $v_{\rm min}$ to $v_{\rm max}$ was
chosen to encompass the \hi\ emission and the UV absorption from the LA. By
varying $v_{\rm min}$ and $v_{\rm max}$ by $\pm$5\kms, we quantify the error on
$N$(\hi), which propagates directly to the abundance measurements.
The magnitude of this error varies from 0.01 to 0.06\,dex for our seven sightlines.\\

(3) {\it Uncertainty due to beam smearing}:
because of the mismatched beam-size between pencil-beam UV observations
and finite-beam radio observations,
a beam-smearing error should be included in abundance
measurements to account for small-scale structure.
In our case, the radio beam-sizes range from
1\arcmin\ (ATCA),
9.1\arcmin\ (GBT),
10.8\arcmin\ (EBHIS),
to 16.1\arcmin\ (GASS)
The magnitude of the beam-smearing effect on abundance measurements
across these angular scales has been shown to be
$\approx$0.10\,dex \citep{Wa01, Fo10}, so we adopt an error on the abundances
of 0.10 dex to account
for this. Nonetheless, small-scale structure on sub-beam scales
is difficult to quantify and cannot be ruled out, particularly given the highly
fragmentary nature of the LA emission.\\

(4) {\it Uncertainty in log\,$U$}: the error in the value of log\,$U$ derived
from the \sit/\siw\ ratio is $\approx$0.1\,dex. This translates to errors in IC(S)
ranging from 0.01 to 0.03, and errors in IC(O) ranging from 0.001 to 0.004 dex.
The errors on IC(O) are an order of magnitude smaller since IC(O) is a flat
function of log\,$U$, whereas IC(S) increases with log\,$U$. \\

\begin{deluxetable*}{llccc ccccc}
\tabletypesize{\footnotesize}
\label{tab:abun}
\tabcolsep=1.0pt
\tablewidth{0pt}
\tablecaption{Ionization Corrections and Corrected Abundances in the Leading Arm}
\tablehead{ & & \multicolumn{3}{c}{\underline{~~~~~~~~~~~~~~~Input to \emph{Cloudy}~~~~~~~~~~~~~~~}} & \multicolumn{3}{c}{\underline{~~~~~~~~~~Output from \emph{Cloudy}~~~~~~~~~~}} & \multicolumn{2}{c}{\underline{Corrected Abundances}}\\
 Target & Region & log\,$N$(\hi) & log\,$n_\gamma$\tm{a} & log\,$\frac{{\rm Si~III}}{{\rm Si~II}}$\tm{b} & log\,$U$\tm{c} & IC(S)\tm{d} & IC(O)\tm{d} & [S/H]\tm{e} & [O/H]\tm{e}\\
 & & (cm$^{-2}$) & (cm$^{-3}$) & & & & & & }
\startdata
\multicolumn{2}{l}{\bf Confirmed LA Sightlines}\\
 CD14-A05           & LA I    & 18.83$\pm$0.10 & $-$5.66 &          $>-0.42$ & $>-$2.9        & $<-0.58$& $-$0.043$\pm$0.004 & $<-0.07$ & $-$0.90$\pm$0.17 \\
 NGC\,3783          & LA II   & 19.92$\pm$0.10 & $-$5.55 &  $-$1.95$\pm$0.15 & $-$3.4$\pm$0.1 & $+$0.003$\pm$0.001 & $-$0.006$\pm$0.001 & $-$0.62$\pm$0.20 & $>-$1.67\tm{f}\\ 
 NGC\,3125          & LA III  & 19.05$\pm$0.05 & $-$5.78 &  $-$0.56$\pm$0.36 & $-$3.0$\pm$0.1 & $-$0.37$\pm$0.02 & $-$0.011$\pm$0.002 & $<$+0.24         & $-$1.40$\pm$0.30\\
 UVQS\,J1016-3150   & LA III  & 18.47$\pm$0.05 & $-$5.73 &  $-$0.16$\pm$0.17 & $-$2.7$\pm$0.1 & $-$0.90$\pm$0.03 & $-$0.014$\pm$0.001 & $<$+0.26         & $-$1.21$\pm$0.33\\
 \hline
\multicolumn{2}{l}{\bf Potential LA Sightlines}\\
 PKS\,1101-325    & LA II/III & 19.30$\pm$0.05 & $-$5.54 &          $>-0.66$ & $>-$3.0 & $<-0.57$ & $-$0.014$\pm$0.001 & $<-0.47$ & $>$15.03 \\ 
 IRAS\,F09539-0439  & LA Ext. & 19.21$\pm$0.05 & $-$5.66 &          $>-0.51$ & $>-$2.9 & $<-0.29$ & $-$0.015$\pm$0.002 & $<$+0.07 & $-$0.99$\pm$0.14\\
 SDSS\,J0959+0503   & LA Ext. & 18.74$\pm$0.10 & $-$5.48 &          $>-0.83$ & $>-$3.2 & $<-0.57$ & $-$0.036$\pm$0.002 & $<$+0.41 & $<-$0.93\\

\vspace{-4mm}
\enddata
\tn{a}{Logarithm of density of ionizing photons in this direction at $d$=20\,kpc given our 3-D radiation field model (see Section 3.1).}
\tn{b}{Logarithm of \sit/\siw\ column-density ratio in the LA component, measured using AOD integrations to \sit\ $\lambda$1206 and \siw\ $\lambda$1193 over the velocity intervals given in Table 2. Limits are 3$\sigma$. These limits propagate to a limit on [S/H].}
\tn{c}{Ionization parameter, $U\equiv n_\gamma/n_{\rm H}$.}
\tn{d}{Ionization corrections for sulfur and oxygen at the inferred value of log\,$U$.}
\tn{e}{Ionization-corrected sulfur and oxygen abundances, calculated by applying ICs to ion abundances presented in Table 2. Errors include statistical errors on the \hi\ and metal column densities, a beam-smearing error of 0.10\,dex, a velocity integration range error, a distance error, and an error on the ionization parameter (see Section 3.2).}
\tn{f}{Lower limit given due to saturation in \oi\ $\lambda$1302. See Richter et al. (2018) for further analysis.}
\end{deluxetable*}



\section{Results}
The oxygen and sulfur abundances in Table~3 
are the principal observational results from this paper.
The most striking result is that the oxygen abundances in the LA are
\emph{low}. We have three good measurements that show this:
[O/H]=$-$0.90$\pm$0.17 (12.6$^{+6.0}_{-4.1}$\% solar) in LA I toward \cds, 
[O/H]=$-$1.40$\pm$0.30 (4.0$^{+4.0}_{-2.0}$\% solar) in LA III toward \ngcb, and 
[O/H]=$-$1.21$\pm$0.33 (6.2$^{+7.0}_{-3.3}$\% solar) in LA III toward \uvqs. 
We also measure a lower limit on the oxygen abundance in LA II
toward \ngca\ of [O/H]$>-1.67$ ($>$2.1\% solar), 
but the \oi\ $\lambda$1302 line is 
clearly saturated so the true oxygen abundance is higher.
Richter et al. (2018, in prep.) find [O/H]=$-$0.54$\pm$0.23 (29$^{+20}_{-12}$\% solar)
in this sightline based on analysis of a high-resolution \hst/STIS E140M
spectrum. Together, these results indicate the oxygen abundances
in the LA vary from 4--29\% solar. 

For the sulfur abundances in the LA, we report one measurement and
three upper limits, although none of the limits are constraining.
The measurement is toward \ngca\ (LA II),
where we derive [S/H]=$-$0.62$\pm$0.16, 
i.e. (S/H)=24$^{+11}_{-7}$\% solar, which is slightly lower than the
published value in this sightline of (S/H)=35$\pm$7\% solar
formed by updating the \citet{Lu98} measurement to the \citet{As09} solar
abundances.
The upper limits on S/H in two cases (toward \ngcb\ and \uvqs) are
due to the non-detection of the \sw\ triplet in the LA component in
these directions. The upper limit toward \cds\ arises for a different reason.
\sw\ $\lambda$1253 absorption at LA velocities is detected in this direction
(1250 and 1259 are both blended), and taken at face value the
\sw\ 1253 line strength would indicate a high sulfur abundance,
[S/H]$\approx-0.05$; however, this is formally an upper limit
because of the uncertainty in IC(S).
Specifically, saturation in the \sit\ 1206 line leads to a lower limit
on the \sit/\siw\ ratio, which propagates to an upper limit on the sulfur
abundance via a limit on the ionization correction.

To further investigate the relationship between \oi, \sw, and \hi,
we show in Figure~3 
the apparent column density profiles
of \oi\ $\lambda$1302, \sw\ $\lambda$1253, and \hi\ 21 cm in all six LA
directions where either \oi\ or \sw\ (or both) is detected.
Of the three lines in the \sw\ triplet, $\lambda$1253 is chosen here since
$\lambda$1259 is blended and $\lambda$1250 is a
weaker line, often undetected. These profiles provide a linear measure of
the absorbing column
in a pixel-by-pixel manner, allowing the distribution of absorbing gas
in velocity space to be analyzed.
In several cases (toward \cds\ and \ngcb)
offsets in velocity centroids of $\approx$10--20\kms\ 
are observed between the metal (\oi\ or \sw) and \hi\ profiles. In other cases
(e.g. \ngca, \uvqs) the metal and \hi\ profiles align closely.
The offsets are likely related to small-scale structure in the gas along the
line of sight, and the beam-smearing error discussed in Section 3.2.
In the case of \cds, the offset may also be related to some of the \hi\
arising behind the star, whereas the absorption by necessity must lie
in front of the star.
These profile differences show
that the assumption the metals and \hi\ are co-spatial (which is implicit
in our \emph{Cloudy} photoionization simulations) is not fully true in
all cases, but our \emph{integrated} metal column densities, from which the
abundances are derived, should be accurate nonetheless.

\begin{figure*}[!ht]
\epsscale{1.0}
\label{fig:nav}
\plotone{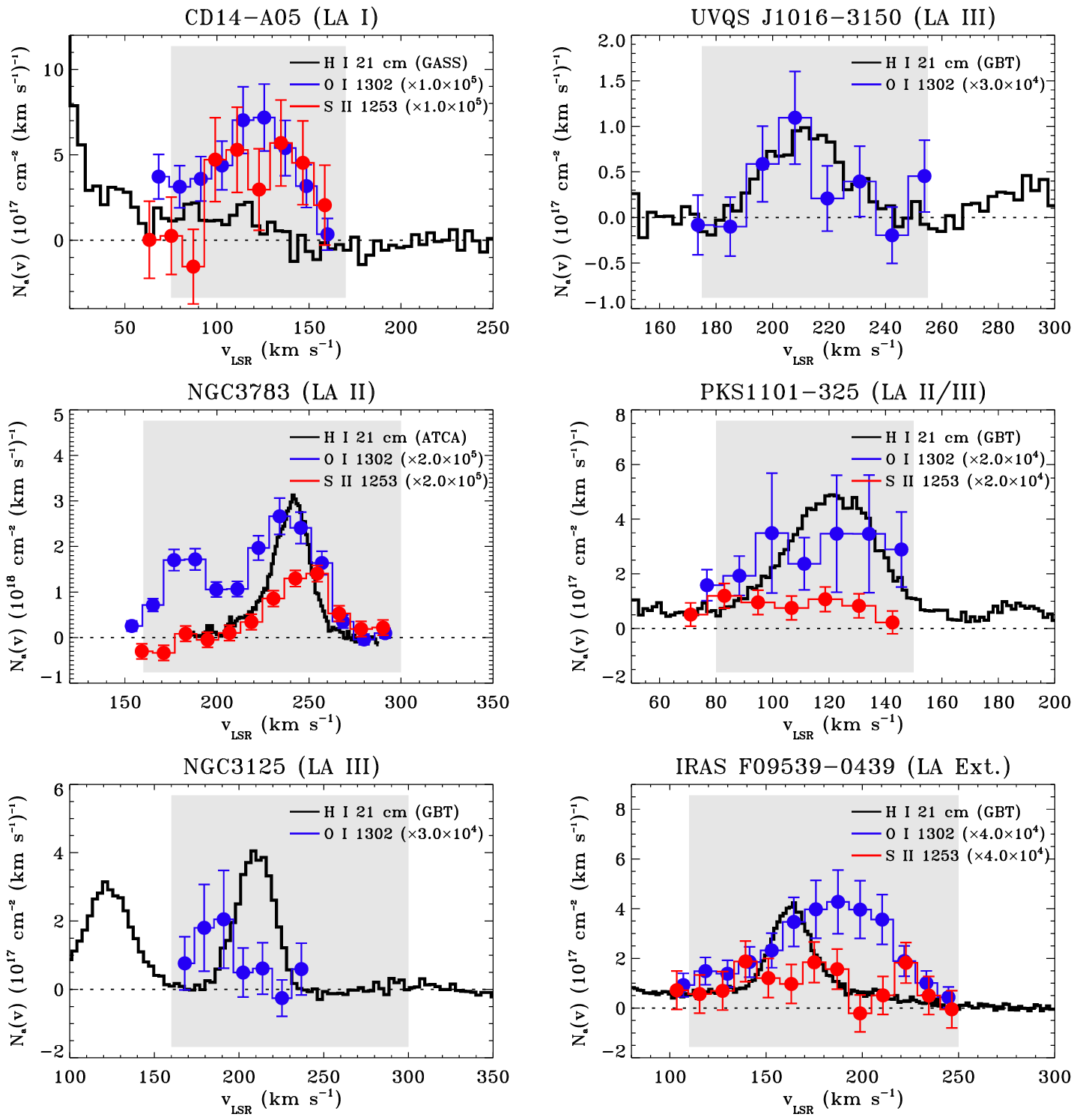}
\caption{Apparent column density profiles per unit velocity
  of \oi\ $\lambda$1302,
  \sw\ $\lambda$1253, and \hi\ 21cm emission in each of the six
  sightlines in our sample where LA metallicity
  measurements are possible (the SDSS J0959+0503 sightline has no \oi\ or \sw\
  detection from the LA so is not shown). The \oi\ and \sw\
  profiles have been rebinned by five pixels and scaled by the
  factors indicated in the legend, for ease of
  comparison. The shaded gray regions show the velocity interval
  over which our LA measurements are made.
  The title indicates in parentheses the LA region probed by the sightline.}
\end{figure*}

\section{Discussion}
\subsection{Source of the LA}
Our analysis is framed by the following questions.
Did the LA originate in the LMC, SMC, or both (like the Magellanic Stream)?
Did it form in a single gas-removal event or in multiple episodes?
And was the event or process that created the Stream the same event or process
that created the LA? The observed chemical enrichment pattern along the
LA directly addresses these questions.
In a single removal event, we expect uniform chemical composition
along the Stream, since once the gas is stripped from the Magellanic Clouds
its chemical abundances are frozen in.
Conversely, in a multiple-removal 
scenario, the gas in different regions would have different metallicities,
with the older regions having lower abundances. In this event, we would
expect the region furthest from the Clouds (LA III) to have lower metallicity
than the region closest to them (LA I).
Any metal mixing between the LA and the surrounding medium would complicate
this picture, since it could dilute or enrich the LA gas over time
\citep{Gr14, He17} depending on the contrast between the LA and the halo
metallicity. This contrast is not well constrained observationally.

With that introduction, our first observation is that
all regions of the LA show metallicities below the current-day metallicity of
the LMC (46\% solar) and the SMC \citep[22\% solar;][]{RD92, Tr07},
with the single exception of the sulfur abundance of 34$\pm$7\% solar 
in the high $N$(\hi) core of LA II.
Our three LA measurements of O/H range from
4.0$^{+4.0}_{-2.0}$\% solar to 12.6$^{+6.2}_{-4.1}$\% solar 
and are therefore
well below $Z_{\rm LMC}$ and $Z_{\rm SMC}$.
This simple point is worth emphasizing, because
it indicates that the LA was formed either (1) long ago (several Gyr ago),
when the LMC and SMC abundances were lower
according to their age-metallicity relations
\citep{PT98, HZ04, HZ09, Me14},
or (2) from the outer regions of the LMC or SMC where the metallicity
is lower than in the inner regions because of radial abundance gradients.
Evidence for radial abundance gradients in the Magellanic Clouds
based on stellar and nebular metallicities is mixed:
\citet{To17} find no strong gradient in \ion{H}{2} region abundances
in either galaxy, and \citet{Ci09}
find no significant gradient in AGB-star abundances in the SMC.
However, \citet{Ci09} report a gradient in AGB-star abundances in the LMC of
$-$0.047$\pm$0.003 dex\,kpc$^{-1}$ out to 8\,kpc,
and \citet{Do14} find a gradient of $-$0.075$\pm$0.011\,dex\,deg$^{-1}$
for red giant stars in the inner 5$\degr$ of the SMC.
Therefore both the LA age and Magellanic abundance gradients may be factors in
explaining the LA's low metallicity.

Our second observation is that there is some evidence for
spatial variation of the oxygen abundances.
The two directions through region LA III have identical (low) oxygen abundances,
given the observational errors, with
[O/H]=$-$1.40$\pm$0.30 toward \ngcb\ and 
[O/H]=$-$1.21$\pm$0.33 toward \uvqs. 
For region LA II, we only report a limit on [O/H], 
but Richter et al. (2018, in prep.) find a higher value of
[O/H]=$-$0.54$\pm$0.23 in this sightline based on
analysis of a high-resolution \hst/STIS spectrum.
Region LA I also has a slightly higher abundance,
[O/H]=$-$0.90$\pm$0.17 (13\% solar), 
based on the measurement toward \cds.
\emph{This observed abundance gradient, with LA III having lower metallicity
  than LA I and II, is consistent with LA III being older than LA I and II},
which is important considering that LA III lies far away from the
Magellanic Clouds on the sky (see Figure 1) and thus has a longer travel
time to reach its current location. The gradient supports a multiple-episode
formation scenario for the LA.

However, the existence of this gradient should be considered tentative
until more data are available, especially considering the statistical
uncertainties. Another caveat is that our single abundance measurement in LA I
(toward \cds) is derived from a stellar sightline, not an AGN sightline,
and we cannot rule out the possibility of some self-enrichment of the LA in
this direction by recent star formation.
Moreover, some of the \hi\ emission toward \cds\ may exist
\emph{behind} the star, whereas the metal absorption lies in front.
This possibility, which is suggested by the mismatch in velocity centroid
between \hi\ emission and UV metal absorption in this direction (Figure 2),
could potentially lead to an underestimate of the LA oxygen abundance,
since it would cause an overestimation of the foreground \hi\ column.

Our third observation is that the very low ($\approx$4--6\% solar) oxygen abundances
measured in region LA III are too low to support an LMC origin.
Even the abundances in regions LA I ($\approx$11\% solar) and LA II ($\approx$29\% solar) 
are difficult to reconcile with a recent LMC origin unless the gas was stripped
from the outer regions of the LMC.
\emph{This strongly suggests that LA III, and potentially all the LA,
is formed from material from the SMC, not the LMC}. 
This stands in contrast to studies of the LA \hi\ kinematics,
which find the LA arises almost entirely from the LMC
\citep{Pu98, Ni08, Ve12}, but it supports several tidal and ram-pressure models
of MS and LA formation \citep{GN96, Co06, DB12, Be12, Ya14}, which predict
the MS and LA form together from SMC material.

Furthermore, the oxygen abundances in the LA match the $\alpha$-element
abundances measured in most of the trailing Magellanic Stream;
oxygen or sulfur abundances of 10\% solar or less have been measured
in seven directions through the SMC filament of the Stream
\citep{Fo10,Fo13,Ku15,Ho17}, though
there is also an LMC filament with 50\% solar abundances \citep{Gi00, Ri13}.
Our LA metallicities can also be compared to those measured in the Magellanic
Bridge, the gaseous connection between the LMC and SMC.
A Bridge metallicity of 10\% solar or lower has been reported
in several studies \citep{Le01, Le02, Mi09} targeting stars
and background quasars. \emph{The finding that the
Bridge, most of the LA, and most of the Stream have low metallicity
($\la$10\% solar) provides support for an SMC origin for most of the
gas in the Magellanic system.}

\subsection{Origin Mechanism of the LA}
The similarity in oxygen abundances between regions LA III
and the SMC filament of the MS (which represents the majority of
the MS) supports an origin model in which LA III and the MS were 
generated in the same event in the past, when the SMC had a lower metallicity.
Most tidal and ram-pressure models of the MS date its
age to $\approx$1.5--2.5\,Gyr \citep{GN96, YN03, Ma05, Co06, Be10, DB12}
although a direct LMC-SMC collision may have happened more recently
\citep[$\sim$100--500\,Myr ago;][]{Be12, Ha15}.
Independently, age-metallicity studies of the stellar population of the
SMC find that it had a mean metallicity of 0.1 solar $\approx$2\,Gyr ago
\citep{PT98, HZ04}. These two findings are consistent
with the idea that much of the Stream and LA were formed at that time
from SMC gas released from an LMC-SMC encounter.

If the LA is at $d$=20 kpc as measured in two separate regions
\citep{MG09, CD14}
then it cannot be a purely tidal feature produced at the first passage of
the LMC and SMC. The reason is that the Clouds are at $\approx$50--60 kpc, so
in a first-passage scenario \citep{Be07, Be10, Be12} \emph{the LA should be
close to or beyond that distance, not 20 kpc}
\citep[e.g. see Figure 10 in][]{Be12}.
Conversely, in a multiple-passage scenario
the LA may have had time to fall down closer to the MW. Our LA abundance
measurements do not resolve this issue but do provide constraints on when
the gas was removed from the Clouds, since they indicate the gas is
chemically unenriched.

In addition to tidal forces and ram pressure, another physical
process relevant to the formation of the MS and LA may be stellar feedback.
The LMC is known to drive an outflow, as seen in UV absorption
lines blueshifted with respect to the LMC systemic velocity
\citep{LH07, Le09, Ba16}, although the LMC outflow does not contain
enough mass flux to reproduce the mass of the Stream \citep{Ba16}.
In the SMC, evidence for stellar feedback was seen in \ion{O}{6}
absorption by \citet{Ho02}, who report \ion{O}{6} column densities correlated
to proximity to star-forming regions,
and \ion{O}{6} kinematics consistent with  a galactic fountain.
These studies indicate that feedback from star formation
drives material out of the Magellanic Clouds into their halos,
where ram-pressure and tidal forces may
disperse it over larger distances \citep{Ol04, Ni08, Ni10, Be12}.
The finding that the LMC filament of the MS can be traced back to
a region near 30 Doradus \citep{Ni08}, an actively star-forming region,
supports this connection between stellar feedback and Stream formation,
although the low LA metallicity we measure does not favor an LMC origin.
Indeed, the lack of evidence for a half-solar LA filament as a counterpart
to the known half-solar MS filament is an important result of our study,
since purely tidal models would be expected to produce such a counterpart.

\subsection{Stellar Abundances in the LA}
For the stellar component of the LA, chemical abundances were measured 
by \citet{Zh17} for a sample of early B-stars located in regions
LA I, II and III discovered by \citet{CD14}.
They found that the five kinematical members of the LA have an average Mg
abundance of [Mg/H]=$-$0.42$\pm$0.16, close to the LMC abundance,
and significantly lower than that that of non-members. However, individually,
our target \cds\ has [Mg/H]=$-$0.57$\pm$0.35,
compatible within errors with the SMC Mg abundance
[Mg/H]$_{\rm SMC}$=$-$0.8$\pm$0.1 \citep{Tr07},
and also compatible with the gas-phase oxygen
abundance we measure in the same sightline, [O/H]$_{\rm gas}$=$-$0.90$\pm$0.17. 
This similarity between the stellar and gas-phase metallicity
is interesting given the possibility that some of
the \hi\ lies behind the star, such that [O/H]$_{\rm gas}$ is formally
a lower limit on the LA oxygen abundance in this direction.
Furthermore, the B-type stars (like \cds) may have formed from gas 
enriched by previous episodes of star formation.
More data are needed to explore the relationship between the stellar
and gas-phase abundances in the LA in more depth. %

\subsection{Outlying Fragments and Size of the LA}
The three sightlines in our sample lying away from the main regions
of the LA are those toward \pks\ (between LA II and LA III),
\irasf\ ($\approx$15$\degr$ north-west of LA III), and 
\sdss\ ($\approx$30$\degr$ north-west of LA III). 
All three directions show HVCs with similar \hi\ columns, with
log\,$N$(\hi) between 18.74 and 19.30. 
Two of the three show kinematics consistent with those of the main
LA regions (the exception is \sdss, where the HVC velocity of 289\kms\
appears too high for an LA origin).
The sulfur and oxygen abundances in these HVCs are given
in the lower part of Table 3.
Toward \pks, \irasf\ and \sdss, we find [O/H]$>-0.98$,
[O/H]=$-$0.99$\pm$0.14 and [O/H]$<-0.93$ (3$\sigma$), respectively. 
These abundances are all close enough to our confirmed LA abundances to suggest
that the HVCs are outlying fragments of the LA: they have the right
kinematics, \hi\ columns, and oxygen abundances to support this idea.
In this interpretation the LA extends $\approx$20$\degr$
further to the north-west (in Galactic coordinates) than
the named regions (LA I, II, and III),
and the outer fragments trace the full extent of the debris field.
The LA then covers a linear extent of $\approx$80\degr\
and an area on the sky of $\approx60\degr\!\times\!80\degr$. 
Nonetheless, this extended Leading Arm is still considerably shorter
than the trailing Stream, which is $\approx$140\degr\ long \citep{Ni10}.
This asymmetry is an important constraint for origin models
\citep[e.g.][]{Pa18}.
The presence of fragmented small-scale structure has long been noticed as a key
feature of the neutral gas in the Magellanic System, including the LA
\citep{St02, Pu03, WK08, Ve12, For13}. Our new results extend the region
over which such small-scale structure is seen.

The finding that the LA is larger than traditionally thought
is reproduced in several simulations \citep[e.g.][]{Ya14, Ha15}.
It is also consistent with results from other regions of the
Magellanic System, where UV absorption detections are reported
in directions up to 30$\degr$ away from the \hi\ Stream
\citep{Fo05, Fo14, Ku15, Ho17}.
Furthermore, \citet{For13} report a bridge of LA cloudlets in-between
LA II and LA III; our HVC at +150\kms\ toward \pks\
(which lies half way between LA II and LA III) may trace this
gaseous bridge. The known presence of LA gas in this region supports
our interpretation of the +150\kms\ HVC as tracing the LA.

\section{Summary}
Using new and archival \hst/COS spectra, we have studied the Leading Arm
in UV absorption of four sightlines, sampling each of the three main regions
LA I, LA II, and LA III. We also present analysis of the HVC absorption
in three additional nearby COS sightlines to investigate whether these HVCs
represent outlying fragments of the LA. We combined the COS data with
publicly available \hi\ 21\,cm spectra and applied a chemical abundance
analysis, focusing on the gas-phase O/H 
ratio because of its small dust correction.
We ran tailored \emph{Cloudy} photoionization simulations
to calculate the ionization corrections. We have arrived at the following
conclusions.

\begin{enumerate}

\item The oxygen abundances in all four sightlines through the LA are
\emph{low}, ranging from
[O/H]=$-$1.40$\pm$0.30 (4.0$^{+2.0}_{-2.0}$\% solar) toward \ngcb\ to
[O/H]=$-$0.90$\pm$0.17 (12.5$^{+6.0}_{-4.1}$\% solar) toward \cds, and
[O/H]=$-$0.54$\pm$0.23 (29$^{+20}_{-12}$\% solar) toward \ngca\ 
(Richter et al. 2018, in prep).
These oxygen abundances are unlikely to be affected by dust or ionization
corrections.

\item We observe a variation in metallicity with location in the LA.
The abundance measurements in region LA I (O/H=12.5\% solar) and
LA II (O/H=25\% solar) are \emph{higher} than the two measurements in
region LA III (O/H=4--6\% solar). 
LA III is the region farthest from the LMC, so its lower metal enrichment
suggests it is the oldest part of the LA.

\item The LA III oxygen abundances are too low to support an LMC
origin and indicate an SMC origin, even when accounting
for the evolution in metallicity of the Magellanic Clouds over the
last few Gyr. The origin of LA I and LA II cannot be conclusively
determined from their oxygen abundances alone, but the observed
abundance gradient is consistent with an SMC origin
for the entire LA.

\item The low oxygen abundances in the LA, particularly in LA III,
match those measured in the SMC filament of the trailing Magellanic Stream.
This is consistent with at least parts of the LA 
being generated in the same event that generated the Stream,
as predicted by many simulations of Stream formation.
Tidal models of the Stream and the age-metallicity
relation of the SMC suggest this event happened $\approx$1.5--2.5\,Gyr ago
via an LMC-SMC close encounter, although a direct collision
may have occurred more recently (100--500 Myr ago).

\item The oxygen abundances, \hi\ columns, and kinematics of the HVCs
measured in two directions (\pks\ and \irasf) 
away from the main LA complexes are consistent with a LA origin.
One of these directions lies $\approx$20$\degr$ further
to the Galactic northwest than LA III. This suggests (but does not prove)
the LA has a larger spatial extent than previously thought, forming an
extended debris field covering a $\approx60\degr\!\times\!80\degr$ region.

\end{enumerate}

In an upcoming paper we will present an in-depth analysis of
the \ngca\ sightline (passing through LA II) using \emph{FUSE} and STIS data
(Richter et al. 2018, in prep). The unusually high
\hi\ column in this direction allows a wide range of molecular and
low-ion metal lines to be detected, enabling a detailed abundance and
physical-conditions analysis beyond the scope of the current paper.

In closing, we note that the chemical abundances presented in this paper,
and specifically the low oxygen abundances indicative of an SMC origin, 
are important clues to constrain numerical simulations
of MS and LA formation.
Purely tidal models for MS predict that trailing and leading arms
form at the same time, but our abundance analysis shows no evidence yet for
a \emph{leading} LMC filament as a counterpart to the
trailing LMC filament. This seemingly indicates that purely tidal models
are unable to explain the LA, and that another physical process such
as ram pressure or stellar feedback may contribute. Further comparisons
between observations and simulations will be necessary to piece together
the full interaction history of our closest satellite neighbors.

\vspace{1cm}
{\it Acknowledgments.}
Support for program 14687 was provided by NASA 
through grants from the Space Telescope Science Institute, which is 
operated by the Association of Universities for Research 
in Astronomy, Inc., under NASA contract NAS~5-26555.
We thank Jerry Kriss for kindly sharing his reductions of the COS data
of \ngca, and we are grateful to the anonymous referee for a useful report.
ED gratefully acknowledges the hospitality of the Center for Computational
Astrophysics at the Flatiron
Institute during the completion of this work.\\

{\it Facilities:} HST (COS), GBT, ATCA, Parkes, Effelsberg.\\





\begin{thebibliography}{}
\bibitem[Anders \& Grevesse(1989)]{AG89} 
Anders, E., \& Grevesse, N. 1989, Geochim. Cosmochim. Acta., 53, 197
  
  
\bibitem[Asplund et al.(2009)]{As09} 
Asplund, M., Greenness, N., Jacques Sauval, A., \& Scott, P. 2009,
\araa, 47, 481

\bibitem[Barger et al.(2013)]{Ba13} 
Barger, K. A., Haffner, L. M., \& Bland-Hawthorn, J. 2013, \apj, 771, 132

\bibitem[Barger et al.(2016)]{Ba16} 
Barger, K. A., Lehner, N., \& Howk, J. C. 2016, \apj, 817, 91




\bibitem[Besla et al.(2007)]{Be07} 
Besla, G., Kallivayalil, N., Hernquist, L. et al. 2007, \apj, 668, 949

\bibitem[Besla et al.(2010)]{Be10} 
Besla, G., Kallivayalil, N., Hernquist, L. et al. 2010, \apj, 721, L97

\bibitem[Besla et al.(2012)]{Be12} 
Besla, G., Kallivayalil, N., Hernquist, L., et al. 2012, \mnras, 421, 2109
  
\bibitem[Bland-Hawthorn \& Maloney(1999)]{BM99} 
Bland-Hawthorn, J., \& Maloney, P. R. 1999, \apj, 510, L33



\bibitem[Boothroyd et al.(2011)]{Bo11}
Boothroyd, A. I., Blagrave, K., Lockman, F. J., et al. 2011, 
\aap, 536, A81

\bibitem[Bordoloi et al.(2017)]{Bo17} 
Bordoloi, R., Fox, A. J., Lockman, F. J., et al. 2017, \apj, 834, 191

\bibitem[Br\"uns et al.(2005)]{Br05} 
Br\"uns, C, Kerp, J., Staveley Smith, L., et al. 2005, \aap, 432, 45

\bibitem[Casetti-Dinescu et al.(2012)]{CD12} 
Casetti-Dinescu, D. I., Vieira, K., Girard, T. M., \& van Altena, W. F.
2012, \apj, 753, 123

\bibitem[Casetti-Dinescu et al.(2014)]{CD14}
Casetti-Dinescu, D. I., Moni Bidin, C., Girard, R. M., et al. 2014, \apj,
784, L37

\bibitem[Cashman et al.(2017)]{Ca17} 
Cashman, F. H., Kulkarni, V. P., Kisielius, R., Ferland, G. J., \&
Bogdanovich, P. 2017, \ap, 230, 8

\bibitem[Cioni(2009)]{Ci09} 
Cioni, M.-R. L., 2009, \aap, 506, 1137
  
\bibitem[Connors et al.(2006)]{Co06}
Connors, T. W., Kawata, D., \& Gibson, B. K. 2006, \mnras, 371, 108


\bibitem[Diaz \& Bekki(2012)]{DB12} 
Diaz, J. D., \& Bekki, K. 2012, \apj, 750, 36

\bibitem[Dobbie et al.(2014)]{Do14} 
Dobbie, P. D., Cole, A. A., Subramaniam, A., \& Keller, S. 2014,
\mnras, 442, 1680

\bibitem[D'Onghia \& Fox(2016)]{DO16}
D'Onghia, E., \& Fox, A. J. 2016, \araa, 54, 363

\bibitem[Ferland et al.(2013)]{Fe13}
Ferland, G. J., Porter, R. L., van Hoof, P. A. M., et al. 
2013, RMxAA, 49, 137

\bibitem[Field \& Steigman(1971)]{FS71} 
Field, G. B., \& Steigman, G. 1971, \apj, 166, 59
  
\bibitem[For et al.(2013)]{For13} 
For, B.-Q., Staveley-Smith, L., \& McClure-Griffiths N. M. 2013, \apj, 764, 74


\bibitem[For et al.(2016)]{For16} 
For, B.-Q., Staveley-Smith, L., McClure-Griffiths, N. M., Westmeier, T.,
\& Bekki, K. 2016, \mnras, 461, 892

\bibitem[Fox et al.(2005)]{Fo05} 
Fox, A. J., Wakker, B. P., Savage, B. D., et al. 2005, \apj, 630, 332

\bibitem[Fox et al.(2010)]{Fo10} 
Fox, A. J., Wakker, B. P., Smoker J. V., et al. 2010, \apj, 718, 1046

\bibitem[Fox et al.(2013)]{Fo13} 
Fox, A. J., Richter, P., Wakker, B. P., et al. 2013, \apj, 772, 110

\bibitem[Fox et al.(2014)]{Fo14} 
Fox, A. J., Wakker, B. P., Barger, K. A., et al. 2014, \apj, 787, 147

\bibitem[Fox et al.(2016)]{Fo16} 
Fox, A. J., Lehner, N., Lockman, F. J., et al. 2016, \apj, 816, L11


\bibitem[Gardiner \& Noguchi(1996)]{GN96} 
Gardiner, L. T., \& Noguchi, M. 1996, \mnras, 278, 191

\bibitem[Gibson et al.(2000)]{Gi00} 
Gibson, B. K., Giroux, M. L., Penton, S. V., et al. 2000, \aj, 120, 1803

\bibitem[Gritton et al.(2014)]{Gr14} 
Gritton, J. A., Shelton, R. L., \& Kyujin, K. 2014,
\apj, 795, 99



\bibitem[Haardt \& Madau(2001)]{HM01} 
Haardt, F. \& Madau, P. 2001, preprint (arXiv:0106018)


\bibitem[Hammer et al.(2015)]{Ha15} 
  Hammer, F., Yang, Y. B., Flores, H., Puech, M., \& Fouquet, S. 2015,
  \apj, 813, 110

\bibitem[Harris \& Zaritsky(2004)]{HZ04} 
Harris, J., \& Zaritsky, D. 2004, \aj, 127, 1532

\bibitem[Harris \& Zaritsky(2009)]{HZ09} 
Harris, J., \& Zaritsky, D. 2009, \aj, 138, 1243



\bibitem[Henley et al.(2017)]{He17} 
Henley, D. B., Gritton, J. A., \& Shelton, R. L. 2017, \apj, 837, 82
  
\bibitem[Hoopes et al.(2002)]{Ho02} 
Hoopes, C. G., Sembach, K. R., Howk, J. C., Savage, B. D., \&
Fullerton, A. W. 2002, \apj, 569, 233
  
\bibitem[Howk et al.(2017)]{Ho17} 
Howk, J. C., Wotta, C. B., Berg, M. A., et al. 2017, \apj, 846, 141

\bibitem[Jenkins(2009)]{Je09} 
Jenkins, E. B. 2009, \apj, 700, 1299

\bibitem[Jensen et al.(2005)]{Je05} 
Jensen, A. G., Rachford, B. L., \& Snow, T. P. 2005, \apj, 619, 891

\bibitem[Kalberla et al.(2005)]{Ka05} 
Kalberla, P. M. W., Burton, W. B., Hartmann, D., et al. 2005, \aap, 440, 775

\bibitem[Kumari et al.(2015)]{Ku15} 
Kumari, N., Fox, A. J., Tumlinson, J., et al. 2015, \apj, 800, 44

\bibitem[Lehner(2002)]{Le02} 
Lehner, N. 2002, \apj, 578, 126

\bibitem[Lehner et al.(2001)]{Le01} 
Lehner, N., Sembach, K. R., Dufton, P. L., Rolleston, W. R. J.,
\& Keenan, F. P. 2001, \apj, 551, 781
  
\bibitem[Lehner \& Howk(2007)]{LH07} 
Lehner, N., \& Howk, J. C. 2007, \mnras, 377, 687

\bibitem[Lehner et al.(2009)]{Le09} 
Lehner, N., Staveley-Smith, L., \& Howk, J. C. 2009, \apj, 702, 940



\bibitem[Lu et al.(1994)]{Lu94} 
Lu, L., Savage, B. D., Sembach, K. R. 1994, \apj, 426, 563

\bibitem[Lu et al.(1998)]{Lu98} 
Lu, L., Savage, B. D., Sembach, K. R., et al. 1998, \aj, 115, 162

\bibitem[Mastropietro et al.(2005)]{Ma05} 
Mastropietro, C., Moore, B., Mayer, L., Wadsley, J., \& Stadel,
J. 2005, \mnras, 363, 509 

\bibitem[Mathewson et al.(1974)]{Ma74}
Mathewson, D. S., Cleary, M. N., \& Murray, J. D. 1974, \apj, 190, 291

\bibitem[McClure-Griffiths et al.(2008)]{MG08} 
McClure-Griffiths, N. M., Staveley-Smith, L,  Lockman, F. J., et al. 
2008, \apj, 673, L143

\bibitem[McClure-Griffiths et al.(2009)]{MG09} 
McClure-Griffiths, N. M., Pisano, D. J., Calabretta, M. R., et al. 
2009, \apjs, 181, 398


\bibitem[Meschin et al.(2014)]{Me14} 
Meschin, I., Gallart, C., Aparicio, A., et al. 2014, \mnras, 438, 1067
 
\bibitem[Misawa et al.(2009)]{Mi09} 
Misawa, T., Charlton, J. C., Kobulnicky, H. A., Wakker, B. P., \& 
Bland-Hawthorn, J. 2009, \apj, 695, 1382
  
\bibitem[Monroe et al.(2016)]{Mo16} 
Monroe, T. R., Prochaska, J. X., Tejos, N., et al. 2016, \aj, 152, 25

\bibitem[Moore \& Davis(1994)]{MD94} 
Moore, B., \& Davis, M. 1994, \mnras, 270, 209

\bibitem[Morton(2003)]{Mo03} 
Morton, D. C. 2003, \apjs, 149, 205

\bibitem[Nidever et al.(2008)]{Ni08} 
Nidever, D. L., Majewski, S. R., \& Burton, W. B. 2008, \apj, 679, 432 (N08)

\bibitem[Nidever et al.(2010)]{Ni10}
Nidever, D. L., Majewski, S. R., \& Burton, W. B. 2010, \apj, 723, 1618

\bibitem[Olano(2004)]{Ol04} 
Olano, C. A. 2004, \aap, 423, 895

\bibitem[Pagel \& Tautvai\v{s}ien\.{e}(1998)]{PT98} 
Pagel, B. E. J., \& Tautvaiv{s}ien\.{e}, G. 1998, \mnras, 299, 535

\bibitem[Pardy et al.(2018)]{Pa18} 
Pardy, S., D'Onghia, E., \& Fox, A. J. 2018, \apj, submitted 

\bibitem[Putman et al.(1998)]{Pu98} 
Putman, M. E., Gibson, B. K., Staveley-Smith, L., et al. 1998, Nature, 394, 752

\bibitem[Putman et al.(2003)]{Pu03} 
Putman, M. E., Staveley-Smith, L., Freeman, K. C., Gibson, B. K., \& 
Barnes, D. G. 2003, \apj, 586, 170

\bibitem[Putman et al.(2012)]{Pu12} 
Putman, M. E., Peek, J. E. G., \& Joung, M. R. 2012, \araa, 50, 491


\bibitem[Richter et al.(2013)]{Ri13} 
Richter, P., Fox, A. J., Wakker, B. P., et al. 2013, \apj, 772, 111

\bibitem[Richter(2017)]{Ri17} 
Richter, P. 2017, in ``Gas Accretion onto Galaxies'', Eds. A. J. Fox
  \& R. Dav\'e, ASSL, 430, 15

\bibitem[Richter et al.(2017)]{Rich17} 
Richter, P., Nuza, S. E., Fox, A. J. et al. 2017, \aap, 607, A48


\bibitem[Russell \& Dopita(1992)]{RD92} 
Russell, S. C., \& Dopita, M. A. 1992, \apj, 384, 508

\bibitem[Savage \& Sembach(1991)]{SS91} 
Savage, B. D., \& Sembach, K. R. 1991, \apj, 379, 245

\bibitem[Sembach et al.(2001)]{Se01} 
Sembach, K. R., Howk, J. C., Savage, B. D., \& Shull, J. M. 2001, \aj, 121, 992


\bibitem[Stanimirovi\'c et al.(2002)]{St02} 
Stanimirovi\'c, S., Dickey, J. M.; Kr\v{c}o, M., \& Brooks, A. M. 2002,
\apj, 576, 773

\bibitem[Toribio San Cipriano et al.(2017)]{To17} 
Toribio San Cipriano, L., Dom\'inguez-Guzm\'an, G., Esteban, C., et al. 
2017, \mnras, 467, 3759

\bibitem[Trundle et al.(2007)]{Tr07} 
Trundle, C., Dufton, P. L., Hunter, I., et al. 2007, \aap, 471, 625
  
\bibitem[Venzmer et al.(2012)]{Ve12} 
Venzmer, M. S., Kerp, J., Kalberla, P. M. W. 2012, \aap, 547, 12

\bibitem[Viegas(1995)]{Vi95} 
Viegas, S. 1995, \mnras, 276, 268

\bibitem[Wakker \& van Woerden(1997)]{WW97} 
Wakker, B. P. \& van Woerden, H. 1997, \araa, 35, 217 

\bibitem[Wakker et al.(2001)]{Wa01} 
Wakker, B. P., Kalberla, P. M. W., van Woerden, H.,
de Boer, K. S., \& Putman, M. E. 2001, \apjs, 136, 537

\bibitem[Wakker et al.(2002)]{Wa02}
Wakker, B. P., Oosterloo, T. A., \& Putman, M. E. 2002, \aj, 123, 1953 
  
\bibitem[Wakker et al.(2015)]{Wa15} 
Wakker, B. P., Hernandez, A. K., French, D., et al. 2015,
\apj, 814, 40 

\bibitem[Wannier et al.(1972)]{Wa72} 
Wannier, P., \& Wrixon, G. T., \& Wilson, R. W. 1972, \aap, 18, 224

\bibitem[Westmeier \& Koribalski(2008)]{WK08} 
Westmeier, T., \& Koribalski, B. S. 2008, \mnras, 388, L29

\bibitem[Winkel et al.(2016)]{Wi16} 
Winkel, B., Kerp, J., Fl\"oer, L, et al. 2016, \aap, 585, A41

\bibitem[Yang et al.(2014)]{Ya14} 
Yang, Y., Hammer, F., Fouquet, S., et al. 2014, \mnras, 442, 2419

\bibitem[Yoshizawa \& Noguchi(2003)]{YN03} 
Yoshizawa, A. M. \& Noguchi, M. 2003, \mnras, 339, 1135

\bibitem[Zhang et al.(2017)]{Zh17} 
Zhang, L., Moni Bidin, C., Casetti-Dinescu, D. I., et al. 2017, \apj, 835, 285

\end{thebibliography}
\end{document}